\documentclass{elsarticle}
\usepackage{epsfig,graphics,amssymb,amsmath,graphicx,indentfirst,subfig,float,appendix,multirow,pifont,color,soul,esint}
\newcommand{\bl}[1]{\boldsymbol{#1}}

\DeclareMathOperator{\tr}{tr}

\newcommand{\ddt}[1]{\frac{{\rm d} #1}{{\rm d} t}}

\renewcommand{\det}[1]{{\rm det}\left({#1} \right)}
\newcommand{\iinf}{\int_{-\infty}^{\infty}}
\newcommand{\dd}{{\rm d}}
\newcommand{\real}[1]{{\rm real}\left({#1}\right)}
\newcommand{\imag}[1]{{\rm imag}\left({#1}\right)}

\usepackage{amsthm}
\makeatletter
\def\els@aparagraph[#1]#2{\elsparagraph[#1]{#2\@addpunct{.}}}
\def\els@bparagraph#1{\elsparagraph*{#1\@addpunct{.}}}
\makeatother

\begin{document}
\begin{frontmatter}

\title{A modal analysis of the behavior of inertial particles in turbulence}

\author[cornell,CTR]{Mahdi Esmaily}
\author[CTR,stanford]{Ali Mani}
\address[cornell]{Sibley School of Mechanical and Aerospace Engineering, Cornell University, Ithaca, NY 14850, USA}
\address[CTR]{Center for Turbulence Research, Stanford University, Stanford, CA 94305, USA}
\address[stanford]{Department of Mechanical Engineering, Stanford University, Stanford, CA 94305, USA}
  
\date{\today} 
\begin{abstract}
The clustering of small heavy inertial particles subjected to Stokes drag in turbulence is known to be minimal at small and large Stokes number and substantial at $\rm St = \mathcal O(1)$. 
This non-monotonic trend, which has been shown computationally and experimentally, is yet to be explained analytically. 
In this study, we obtain an analytical expression for the Lyapunov exponents that quantitatively predicts this trend. 
The sum of the exponents, which is the normalized rate of change of the signed-volume of a small cloud of particles, is correctly predicted to be negative and positive at small and large Stokes numbers, respectively, asymptoting to $\tau Q$ as $\tau \to 0$ and $\tau^{-1/2} |Q|^{1/4}$ as $\tau \to \infty$, where $\tau$ is the particle relaxation time and $Q(\tau)$ is the difference between the norm of the rotation- and strain-rate tensors computed along the particle trajectory. 
Additionally, the trajectory crossing is predicted only in hyperbolic flows where $Q<0$ for sufficiently inertial particles with a $\tau$ that scales with $|Q|^{-1/2}$.
Following the onset of crossovers, a transition from clustering to dispersion is predicted correctly.
We show these behaviors are not unique to three-dimensional isotropic turbulence and can be reproduced closely by a one-dimensional mono-harmonic flow, which appears as a fundamental canonical problem in the study of particle clustering. 
Analysis of this one-dimensional canonical flow shows that the rate of clustering, quantified as the product of the Lyapunov exponent and particle relaxation time, is bounded by $-1/2$, behaving with extreme nonlinearity in the hyperbolic flows and always remaining positive in the elliptic flows. 
These findings, which are stemmed from our analysis, are corroborated by the direct numerical simulations. 

\end{abstract}
\end{frontmatter}

\section{Introduction} \label{intro}
The characterization, prediction, and design of a wide range of applications rely on understanding the dynamics of inertial particles in turbulent flows in general and particle clustering or dispersion in particular.
The formation of planets and planetesimals in our early solar system is hypothesized to be a result of particle clustering \cite{bracco1999particle, cuzzi2001size, johansen2004simulations}. 
The accurate prediction of weather relies on proper modeling of droplet coalescence in the clouds, a phenomenon with a direct connection to the problem of particle clustering \cite{shaw1998preferential, pinsky2000stochastic, falkovich2002acceleration}. 
The design of a particle-based solar receiver intended for the endothermic chemical reaction with a high-temperature requirement can be achieved through particles clustering \cite{pouransari2017effects, farbar2016monte, zamansky2014radiation, frankel2016settling}.
Apart from the few preceding examples, numerous areas in science and engineering, ranging from sediment and pollutant transport in environmental flows to pharmaceutical applications, also benefit from the better understanding of particle clustering and dispersion \cite{eaton1994preferential, balachandar2010turbulent}.

In a physically realistic scenario, particle clustering is governed by multiple non-dimensional parameters such as the density ratio of particle to carrier flow, the particle volume fraction, the size of particles relative to the characteristic length scale of the flow, the Froude, Stokes, Reynolds, and Knudsen numbers, the effect of the particle shape as well as the anisotropy of the underlying flow.
While a wide range of behaviors is observed in different regions of this multi-dimensional parameter space, particle clustering is primarily governed by a small subset of parameters.
To make the problem analytically tractable, we only focus on that small subset of parameters and neglect the effect of other parameters that are of secondary importance.
Namely, we neglect the effect of the particles on the flow as well as particle-particle interactions, both of which imply low volume and mass fractions.
The effect of thermal fluctuations and body forces on the particle motion are neglected considering regimes with low Knudsen and high Froude numbers, respectively.
In our derivation, we will be considering statistically isotropic flows.
The finite-size effects are neglected by adopting a point-particle approach, which is to consider only particles that are much smaller than the smallest structure of the flow in the absence of particles.
More specifically, we focus on regimes with $d_{\rm p}/\eta \ll 1$, where $d_{\rm p}$ is the particle diameter and $\eta$ is a length scale of the flow that characterizes curvature of the velocity field.
Additionally, we assume that the particle Reynolds number is sufficiently small and that the drag is linearly proportional to the slip velocity. 
We also focus on the regimes at which the particles are much denser than the fluid, i.e. $\rho_{\rm p}/\rho_{\rm f} \gg 1$ with $\rho_{\rm p}$ and $\rho_{\rm f}$ denoting the particle density and the fluid density, respectively.

In such a simplified regime, the particle clustering becomes a function of the particle Stokes number primarily.
The Stokes number represents the inertia of the particles and is defined as the ratio of the particle relaxation time $\tau := \rho_{\rm p} d_{\rm p}^2/(18\rho_{\rm f} \nu)$, where $\nu$ is the fluid kinematic viscosity, to a flow time scale.
For homogeneous isotropic turbulent flows, the flow time scale is often taken to be the Kolmogorov time scale $\tau_{\eta} := \sqrt{(\nu/\epsilon)}$, in which $\epsilon$ is the mean volumetric dissipation rate, respectively.
Since the Kolmogorov length scale $\eta:=(\nu^3/\epsilon)^{1/4}$ is also used for the flow length scale, the Stokes number can be defined as 
\begin{equation}
{\rm St} := \frac{\tau}{\tau_\eta} = \frac{1}{18} \left(\frac{\rho_{\rm p}}{\rho_{\rm f}}\right) \left(\frac{d_{\rm p}}{\eta}\right)^2,
\label{stokes_def}
\end{equation}
which depends on the density ratio $(\rho_{\rm p}/\rho_{\rm f})$ and the length scale ratio $(d_{\rm p}/\eta)$. 
For the regime under consideration, $(\rho_{\rm p}/\rho_{\rm f})$ is large and $(d_{\rm p}/\eta)$ is small, thus St is finite and can vary from 0 to $\infty$, a range that is investigated in this study.
In practice, this regime translates to gas flows laden with small dense droplets or solid particles \cite{pouransari2017effects,farbar2016monte}.

In addition to the Stokes number, the underlying flow must be fully defined for the characterization of particle clustering.
Among the wide range of possibilities, certain canonical flows are studied the most for their physical relevance or their fundamental importance. 
If we limit this investigation to stationary isotropic turbulence as a particular class of flows, the flow Reynolds number will appear as an additional non-dimensional parameter that affects particle clustering.
The Reynolds number determines the range of time scales that a particle encounters along its trajectory.

Particle clustering, as a function of the particle Stokes number and the flow Reynolds number, has been studied in the past extensively. 
It has been observed experimentally \cite{fessler1994preferential, aliseda2002effect, salazar2008experimental, saw2008inertial, eaton1994preferential, petersen2019experimental}, simulated numerically \cite{squires1991preferential, ray2011preferential, calzavarini2008quantifying, tagawa2012three, goto2008sweep, sundaram1997collision}, and described analytically \cite{maxey1987gravitational, robinson1956motion, bec2004multifractal, bec2003fractal, balkovsky2001intermittent}.
These studies have shown that the degree to which particles clustering mainly depends on the Stokes number.
Depending on their Stokes number, particles may homogeneously disperse in space or preferentially concentrate in certain regions of the flow. 
For ${\rm St} \ll 1$, particles become neutral fluid tracers and experience minimum clustering. 
There is also minimal clustering at the limit of St $\gg 1$, where particles follow a ballistic trajectory uncorrelated with the underlying flow. 
The maximum clustering is achieved when ${\rm St} \approx 1$.
This non-monotonic variation, which has been observed experimentally and numerically, is yet to be described analytically.
In this study, we attempt to describe this trend using analytical tools while employing numerical simulations for verification purposes. 

The particle motion subjected to the preceding simplifications is well characterized by the Stokes drag.
Denoting the position of a particle by $\bl x(t)$ and the flow velocity at the particle location by $\bl u(\bl x,t)$, the dimensionless equation of the motion of a particle subjected to the Stokes drag is
\begin{equation}
\ddot {\bl x} = \bl u(\bl x, t)  - \dot{\bl  x},
\label{stokes}
\end{equation}
where $\dot{(\bullet)} := {\rm d} (\bullet) / {\rm d} t$.
We employed $\tau$, $L$, and $U = L/\tau$ as the time, length, and velocity scales to non-dimensionalize corresponding parameters in Eq.~\eqref{stokes}.
All the following equations are also non-dimensionalized based on $\tau$ and $L$. 
$\tau$ is the particle relaxation time, defined above, and $L$ is a characteristic length scale.
The following formulations are independent of the choice of $L$, hence its choice is arbitrary.
Parameters normalized based on the flow time scale rather than the particle relaxation time are distinguished by the subscript $\eta$.
The only exception is $\tau_\eta$ which denotes the dimensional Kolmogorov time scale rather than the particle relaxation time normalized by the flow time scale. 
Although Eq.~\eqref{stokes} appears to be a simple ordinary differential equation, it behaves nonlinearly due to the dependence of $\bl u$ on $\bl x$.
It is this nonlinear behavior that gives rise to particle clustering, which is a complex and nonlinear phenomenon. 

One of the first analytical relations for quantifying particle clustering was derived in \cite{robinson1956motion} and \cite{maxey1987gravitational}. 
This relation, which hereafter is referred to as RM, is obtained by approximating the acceleration of particles with that of the fluid. 
One of the underlying assumptions of this analysis is 
\begin{equation}
\overline{\nabla \cdot\bl u} = 0,
\label{continuity}
\end{equation}
where $\overline{(\bullet)}$ denotes Lagrangian time-averaging along the trajectory of a particle.
This condition, which is less restrictive than $\nabla \cdot \bl u=0$, only requires the flow to be incompressible along particle trajectory over a long period.
Thus, this condition can also be satisfied for compressible flows where the total mass of a given control volume remains the same over a long time interval. 
The central assumption of RM, however, is approximating the acceleration of particles with that of the flow via $\ddot{\bl x} \approx {\rm D}\bl u/{\rm D} t$.
With these two assumptions, one can take the divergence of Eq.~\eqref{stokes} to obtain
\begin{equation}
\mathcal C({\rm St}) := \overline{ \nabla \cdot \dot {\bl x} } \approx - \overline{\frac{\partial u_i}{\partial x_j}\frac{\partial u_j}{\partial x_i}} = \overline{ \|\bl \Omega\|^2 - \|\bl S\|^2} = Q({\rm St}),
\label{RM}
\end{equation}
where $\bl \Omega$ and $\bl S$ are the dimensionless rotation-rate and strain-rate and $Q$ is half of the Q-criterion, which is defined in the literature for identification of the vortical regions \cite{hunt1988eddies, dubief2000coherent}.
The sign of $Q$ indicates the dominance of flow rotation-rate ($Q > 0$) or strain-rate ($Q < 0$), which are associated with elliptic and hyperbolic regions of the flow, respectively.
Apart from the fact that $Q$ is normalized by $\tau$, its computation along the particle trajectory implies its dependence on St, i.e. $Q_\eta = Q_\eta({\rm St})$.
The RM expression relates the degree of preferential concentration, here defined as $\mathcal C$, to the divergence of the particle velocity field. 
The sign of $\mathcal C$ determines the regime of particle clustering or dispersion. 
Dispersion occurs when $\mathcal C > 0$, i.e. when particles move away from each other over time.
Clustering occurs when $\mathcal C < 0$, i.e. when particles get closer to each other over time.
The Eulerian definition of $\mathcal C$ through divergence operation in Eq.~\eqref{RM} is harder to interpret when particle trajectories cross in which $\dot {\bl x}$ is not a well-defined function of the spatial co-ordinates \cite{bec2003fractal}.
Thus, in Section \ref{ana_der}, we adopt a Lagrangian definition of $\mathcal C$, which mathematically corresponds to the above Eulerian definition, for quantification of the particle clustering. 

Approximating the acceleration of particles with that of flow, i.e. $\ddot{\bl x} \approx {\rm D}\bl u/{\rm D} t$, limits the validity of RM to ${\rm St} \ll 1$. 
The accuracy of RM in predicting the first- and second-order statistics of $\mathcal C$ has been shown for ${\rm St} \ll 1$ in homogeneous turbulence \cite{ferry2003locally, esmaily2016CPT} as well as synthetic flows \cite{ijzermans2010segregation}. 
For St $\ge 1$, RM predicts an unbounded $\mathcal C$ proportional to St and fails to capture the non-monotonic behavior of $\mathcal C({\rm St})$.
Additionally, Eq.~\eqref{RM} suggests that particles are repelled from the rotation-dominated regions, where $\|\bl \Omega\| > \|\bl S\|$, and preferentially concentrate in regions with higher strain-rate.
These generally accepted qualitative assessments will be examined thoroughly in this study. 

In an earlier attempt, we derived an alternative relationship for $\mathcal C$ via a first-order correction to RM \cite{esmaily2016CPT}. 
In that study, we linearized Eq.~\eqref{stokes} by expanding $\bl u(\bl x,t)$ using the Taylor series in spatial co-ordinates and expressing its temporal variation in the Fourier space to obtain an eigenvalue problem for the Lyapunov exponents of pairs of inertial particles. 
We showed that the sum of these exponents (in three dimensions $\lambda_1 + \lambda_2 + \lambda_3$) is equal to the divergence of the particle velocity field and can be expressed as
\begin{equation}
\mathcal C({\rm St}) = \iinf \frac{\tilde \rho^{\rm Q}(\omega; {\rm St})}{1 + \omega^2} \dd \omega,
\label{SL}
\end{equation}
where
\begin{equation}
\tilde \rho^{\rm Q}(\omega; {\rm St}) := \frac{1}{2\pi}\iinf \rho^{\rm Q}(t; {\rm St}) e^{-\hat i \omega t} \dd t = \frac{1}{2\pi}\iinf -\overline{\frac{\partial u_i(t^\prime)}{\partial x_j}\frac{\partial u_j(t^\prime+t)}{\partial x_i}} e^{-\hat i \omega t}\dd t.
\label{Q_def}
\end{equation}
In the last integral $t^\prime$ varies along the particle trajectory, thus $\overline{(\bullet)}$ operates on $t^\prime$.
This definition implies $\rho^{\rm Q}(t) = \rho^{\rm \Omega}(t) - \rho^{\rm S}(t)$, where $\rho^{\rm \Omega}$ and $\rho^{\rm S}$ are the norm of the autocorrelation function of the rotation-rate and strain-rate tensors, respectively, and $\tilde \rho^{\rm Q}(\omega)$ is the Fourier transformation of $\rho^{\rm Q}(t)$.

The primary assumption associated with Eq.~\eqref{SL}, which we refer to as SL (Small Lyapunov) hereafter, is that the Lyapunov exponents $\lambda_i$ are much smaller than 1. 
This assumption is valid at relatively small St in which $\lambda_i$ are also small. 
Comparing Eqs.~\eqref{RM} and \eqref{SL}, SL can be considered as the filtered RM which accounts for unresponsiveness of particles to high-frequency fluctuations. 
As St $\to 0$, SL exactly reproduces RM.
At this limit, $\omega^2$ in the denominator of Eq.~\eqref{SL} can be neglected, thus reproducing Eq.~\eqref{RM} exactly. 
Both SL and RM are a linear function of the rotation-rate and strain-rate tensors, predicting a $\mathcal C$ that changes linearly with $Q$.
Due to this linearity, both formulations predict dispersion ($\mathcal C > 0$) only for rotating flows ($Q > 0$) and clustering ($\mathcal C < 0$) only for straining flows ($Q < 0$). 
Neither of them predicts clustering for a rotating flow or dispersion for a straining flow. 
While these predictions agree with the numerical simulations at small St, they break down at higher Stokes numbers. 
Although SL remains bounded and provides a better prediction than RM, it still deviates from the reference for ${\rm St} > \mathcal O(1)$ due to the underlying assumption of $|\lambda_i| \ll 1$. 

The objective of this study is to derive an analytical relationship for $\mathcal C({\rm St}; \tilde \rho^{\rm Q})$ which is valid at both small and large St. 
In what follows, we present a step-by-step derivation of a solution that is extracted from Eq.~\eqref{stokes} for particles experiencing statistically isotropic flow.
Then, we compare the prediction of RM, SL, and our solution to the reference numerical results. 
For this purpose, we first consider a one-dimensional unimodal oscillatory flow.
Through this case, we test the accuracy of the proposed solution and discuss its implications. 
Then, we extend our analysis to more complex multi-dimensional flows, oscillating at a continuous range of frequencies. 
We consider three-dimensional isotropic turbulence for this purpose, where we compare all the analyses at a wide range of Stokes numbers.

\section{Analytical derivation} \label{ana_der}
\begin{figure}
\begin{center}
\includegraphics[width=0.6\textwidth]{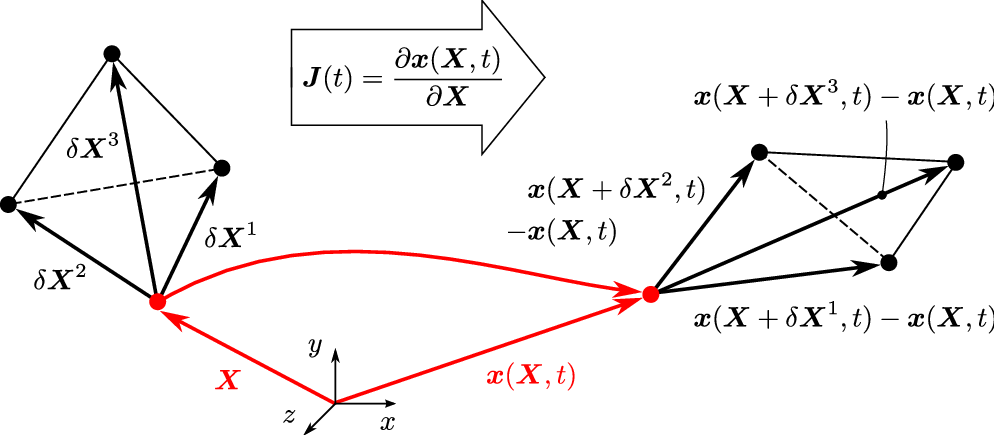}
\caption{A collection of four particles initially located at $\bl X$ and $\bl X+\delta \bl X^i$, $i=1,2,3$, move in time to positions $\bl x(\bl X,t)$ and $\bl x(\bl X + \delta \bl X^i,t)$, respectively.
The relative motion of particles can be characterized by the deformation tensor $\bl J(t)$.}
\label{fig:transformation}
\end{center}
\end{figure}

Consider a collection of four particles that are located at $\bl X$ and $\bl X+\delta \bl X^i$, $i=1,2,3$, at $t=0$ (Figure \ref{fig:transformation}).
We refer to this collection of inertial particles, which are within an infinitesimal distance from each other, a cloud.
At time $t$, the particles in the cloud will move to new positions $\bl x(\bl X,t)$ and $\bl x(\bl X+\delta \bl X^i,t)$, respectively.
Without loss of generality, we choose $\delta$ such that the particles are always at relatively small distances from each other.
More specifically, $\|\bl x(\bl X+\delta \bl X^i,t) - \bl x(\bl X,t) \| \ll \eta$ is a sufficient condition to ensure particles in the cloud experience a linearly-varying velocity field.
Given this condition, the trajectory of an arbitrary particle within the cloud that is initially located at $\bl X + \delta \bl X$ can be described as
\begin{equation}
x_i(\bl X + \delta \bl X, t) = x_i(\bl X, t) + \frac{\partial x_i}{\partial X_j} \delta X_j.
\label{x_def}
\end{equation}
One may choose to consider a cloud as a collection of a larger number of particles than the one that is shown in Figure \ref{fig:transformation}.
Nevertheless, based on Eq.~\eqref{x_def}, the motion of all those particles can be represented as the linear combination of the motion of $n$ particles in an $n$-dimensional space.
In other words, the motion of a cloud is fully characterized via Eq. \eqref{x_def} once $\bl x(\bl X,t)$ and the linear deformation tensor
\begin{equation}
J_{ij}(t) := \frac{\partial x_i(t)}{\partial X_j},
\label{J_def}
\end{equation}
are defined.

Considering the configuration of the cloud that is fully characterized based on $x_i(\bl X + \delta \bl X^j,t) - x_i(\bl X,t)$ (Figure \ref{fig:transformation}), we define
\begin{equation}
V(t) := \det{ x_i(\bl X + \delta \bl X^j,t) - x_i(\bl X,t) },
\label{vol_def}
\end{equation}
as the signed-volume of the cloud at any given time $t$, where $\det{\bullet}$ is the determinant operator.
Since $\bl x(\bl X,0) = \bl X$, it directly follows from Eq.~\eqref{vol_def} that $V(0) = \det{\delta X^j_i}$.
It is evident from the geometrical relationship between the volume of a tetrahedral and the determinant of its corresponding matrix, the actual volume of the cloud is $|V(t)|/6$, where factor of $1/6$ is specific to a tetrahedral cloud and eliminated from our definition for the sake of notation brevity. 
The sign of $V(t)$, on the other hand, denotes whether particles are oriented relative to each other following a right-hand rule. 
We choose the initial orientation of particles such that always $V(0) > 0$.
Thus, $V(t)<0$ only occurs when one of the particles passes through the plane constructed by the other three (or the line constructed by the other two in two dimensions or crossing over the other particle in one dimension), turning the cloud inside-out. 
Hereafter, we refer to such incidents as particle trajectory crossing.

To express $V(t)$ in terms of $\bl J$, we rearrange Eq.~\eqref{x_def} for a collection of particles and use Eq.~\eqref{J_def} to obtain
\begin{equation}
x_i(\bl X + \delta \bl X^j, t) - x_i(\bl X, t) =  J_{ik} \delta X_k^j.
\label{xj_rel}
\end{equation}
Taking the determinant of Eq.~\eqref{xj_rel} and substituting for $V$ based on Eq.~\eqref{vol_def} yields
\begin{equation}
V(t) =  V(0) \det{\bl J(t)},
\label{vj_rel}
\end{equation}
which shows that the determinant of $\bl J(t)$ is the signed-volume of the cloud, which has undergone a linear deformation characterized by $\bl J$, normalized by its initial volume.
Although we derived Eq.~\eqref{vj_rel} for a tetrahedral cloud composed of four particles (Figure \ref{fig:transformation}), the result is generalizable to a cloud with an arbitrary number of particles by considering the entire cloud as a collection of tetrahedron elements and repeating the above procedure for each element. 

Based one the definition of $V(t)$, we define the finite-time exponential rate of change of the signed-volume, $\mathcal C^t$, such that
\begin{equation}
V(t) = V(0) \exp(\mathcal C^t t),
\label{vc_rel}
\end{equation}
holds. 
Combining Eqs.~\eqref{vj_rel} and \eqref{vc_rel} yields
\begin{equation}
\mathcal C^t := \frac{1}{t} \ln\left[ \det{\bl J(t)}\right],
\label{Ct_def}
\end{equation}
which is an alternative definition of $\mathcal C^t$. 
As shown in \ref{der_C_Ct}, $\mathcal C^t$ is related to $\mathcal C$, which was defined earlier as the divergence of particle velocity field, by
\begin{equation}
\mathcal C = \lim_{t\to \infty} \mathcal C^t.
\label{C_def}
\end{equation}

Before proceeding further, it is necessary to build a physical intuition around $\mathcal C$.
From a Lagrangian perspective, $\mathcal C$ is the exponential rate at which $V(t)$ changes over a long period.
Its real component, which only depends on $|V(t)|$, has an intuitive meaning.
It is the exponential rate of change of the volume of the region in the flow that is contaminated by particles. 
Thus, $\real{\mathcal C} > 0$ signifies the regimes of particle dispersion, which is analogous to diffusion induced mixing in continuum flows.
$\real{\mathcal C} < 0$, on the other hand, signifies the regimes of particle clustering in which the collection of particles come together, causing the cloud of particles to form fractals with a dimension smaller than the number of spatial dimensions.
From Eq.~\eqref{vc_rel}, having a $\mathcal C$ with a nonzero imaginary component indicates the oscillatory behavior of $V(t)$.
It is the particle trajectory crossing that changes the sign of $V(t)$, leading to $\imag{\mathcal C} \ne 0$.
According to Eq.~\eqref{C_def}, $\mathcal C^t$ is closely related to $\mathcal C$.
$\mathcal C^t$ is to $\mathcal C$ what the finite-time Lyapunov is to the Lyapunov exponent. 
By tracking $\mathcal C^t$ in time, one may identify incidents of trajectory crossing at which $V(t) \to 0$ and $\real{\mathcal C^t} \to -\infty$.
We will use these concepts as we develop an expression for $\mathcal C$ for an arbitrary flow and $\rm St$.

Based on Eqs.~\eqref{Ct_def} and \eqref{C_def}, $\mathcal C$ can be computed once $\bl J$ is determined.
To obtain a relationship for $\bl J$, we need the governing equation of motion of particles that is taken to be Eq.~\eqref{stokes}, when the assumptions enumerated in Section~\ref{intro} are considered.
Taking the derivative of Eq.~\eqref{stokes} with respect to $\bl X$ and employing the chain rule, yields
\begin{equation}
\ddot J_{ij} + \dot J_{ij} = \frac{\partial u_i}{\partial x_k} J_{kj},
\label{J_stokes}
\end{equation}
where $\dot J_{ij} = \partial \dot x_i/\partial X_j$ and $\ddot J_{ij} = \partial \ddot x_i/\partial X_j$, which hold true since $\bl X \ne \bl X(t)$. 
Also, $\nabla \bl u = \partial \bl u/\partial \bl x$ is expressed in terms of $\bl x$ and thus is tractable by computing the fluid velocity gradient along the trajectory of the cloud. 
The tensor $\nabla \bl u$ is a general function of time and as a result, Eq.~\eqref{J_stokes} is not a constant coefficient ordinary differential equation (ODE) to be integrated directly. 
Nevertheless, analogous to an ODE with constant coefficients, its solution $\det{\bl J}$ exponentially grows or decays indefinitely in time for $\mathcal C \ne 0$.
To proceed further, we make use of 
\begin{equation}
F_{ij} := \dot J_{ik} J_{kj}^{-1},
\label{F_def}
\end{equation}
a transformation that produces a more tractable constant coefficient ODE.
Note $J_{kj}^{-1} = (\bl J^{-1})_{kj}$ is implied in Eq. \eqref{F_def} and what follows. 
Based on this transformation,
\begin{equation}
\dot F_{ij} = \ddot J_{ik} J^{-1}_{kj} - F_{ik}F_{kj}
\label{F_J_rel}
\end{equation}
and hence from Eq.~\eqref{J_stokes}
\begin{equation}
\dot F_{ij} + F_{ik}F_{kj} + F_{ij} = \frac{\partial u_i}{\partial x_j},
\label{F_eq}
\end{equation}
which has been derived by others as well \cite{falkovich2002acceleration, gustavsson2016statistical} and classifies as the Riccati equation in a tensorial form.

Equation \eqref{F_eq}, in contrast to Eq.~\eqref{J_stokes}, is nonlinear but has constant coefficients. 
It is expressed in terms of $\bl F$, which is an instantaneous rate of deformation of the cloud normalized by its size and is independent of the arbitrary choice of $\bl X$. 
As a result, its determinant $\det{\bl F}$ is a statistically stationary variable for sufficiently long $t$ for flows reaching an equilibrium.
To show that and establish a relationship between $\mathcal C$ and $\bl F$, we employ Jacobi's formula,
\begin{equation}
\tr(\bl F) = \det{\bl J}^{-1} \ddt {\left[ \det{\bl J} \right] },
\label{Jacobi_F}
\end{equation}
along with Eqs.~\eqref{Ct_def} and \eqref{C_def} as outlined in \ref{app:eqC} to obtain
\begin{equation}
\mathcal C = \overline{\tr(\bl F)},
\label{C_cal}
\end{equation}
in which $\rm \tr(\bullet)$ is the trace operator.
This simple relationship indicates that $\mathcal C$, which is the sum of the Lyapunov exponents, is the time average of the sum of the eigenvalues of $\bl F$.
In other words, the eigenvalues of $\overline{\bl F}$ are the Lyapunov exponents associated with the pairs of inertial particles.

Next, we solve Eq.~\eqref{F_eq} for $\bl F$ to find an analytical estimate for $\mathcal C$ via Eq.~\eqref{C_cal}.
Given that the tensor $\nabla \bl u$ is a general function of time, we find a solution for Eq.~\eqref{F_eq} by expressing $\nabla \bl u$ as a set of harmonic functions using the Fourier transformation.
Thus, Eq.~\eqref{F_eq} can be written as
\begin{equation}
\dot F_{ij} + F_{ik}F_{kj} + F_{ij} = \sum_\omega G_{ij}(\omega) e^{\hat i \omega t},
\label{F_om}
\end{equation}
where
\begin{equation}
\bl G(\omega) := \frac{1}{2\pi}\iinf \nabla \bl u(t) e^{-\hat i \omega t} \dd t.
\label{G_def}
\end{equation}
In this equation and all that follows, $\omega$, which runs form $-\infty$ to $\infty$, can be considered as $2\pi k/T$ with $k\in \mathbb Z$ and $T \to \infty$ being the sampling period.
We also express $\bl F$ by taking its Fourier transformation as
\begin{equation}
\bl F = \lambda \bl I + \sum_\omega \bl \Psi(\omega) e^{\hat i \omega t},
\label{F_guess}
\end{equation}
where $\bl I$ is the identity tensor, $\lambda$ is a representative Lyapunov exponent, and tensor $\bl \Psi(\omega)$ is the oscillatory response of $\bl F$ to $\bl G$. 
In Eq.~\eqref{F_guess}, the steady response, which is expressed as $\lambda \bl I$, is separated from $\bl \Psi$ to simplify the form of following expressions.
The form of the steady term implies that the Lyapunov exponents are equal, which need not be true in general.
The problems considered in this study, where the flow around the particles is assumed to be statistically isotropic, have no preferred directions.
For such problems, all exponents are equal\footnote{Distinct $\lambda_i$ are reported in the literature even for isotropic flows, which are obtained by sorting each ensemble before averaging. If $\lambda_i$ correspondence with $i^{\rm th}$ co-ordinate is preserved -- statistical uncertainty associated with the finite sampling aside -- then the ensemble averages in all directions will be equal owing to the isotropy condition.}, and the form of the steady term in Eq.~\eqref{F_guess} is valid.
Also, note that from Eqs.~\eqref{C_cal} and \eqref{F_guess} $\mathcal C$ can be computed as
\begin{equation}
\mathcal C = n \lambda,
\label{C_lambda}
\end{equation}
where $n$ is the number of spatial dimensions.

Our goal is to determine $\lambda$ and $\bl \Psi$ using Eq.~\eqref{F_om}.
Substituting for $\bl F$ in Eq.~\eqref{F_om} yields
\begin{equation}
(\lambda + \lambda^2) I_{ij} + \sum_\omega (1 + 2\lambda + \hat i \omega)\Psi_{ij}(\omega) e^{\hat i \omega t} + \sum_{\omega_n}\sum_{\omega_m} \Psi_{ik}(\omega_n) \Psi_{kj}(\omega_m) e^{\hat i (\omega_n + \omega_m) t} =  \sum_\omega G_{ij}(\omega) e^{\hat i \omega t},
\label{F_1_i}
\end{equation}
To proceed further, we neglect the higher-order terms in Eq.~\eqref{F_1_i} by assuming $\|\bl \Psi\| \ll 1$.
We revisit this assumption in \ref{app:HOE}, where the effect of higher order terms is analyzed for a unimodal excitation that is a model problem introduced in Section~\ref{unimodal}. 
Based on this assumption, all the time-dependent terms in the second summation can be neglected compared to those in the first summation. 
The remaining time-independent terms are retained as they might be comparable to the first term.
Thus, keeping only the terms with $\omega_n = -\omega_m$ yields
\begin{equation}
(\lambda + \lambda^2) I_{ij} + \sum_\omega (1 + 2\lambda + \hat i \omega) \Psi_{ij}(\omega) e^{\hat i \omega t} + \sum_\omega \Psi_{ik}(\omega) \Psi_{kj}(-\omega) = \sum_\omega G_{ij}(\omega) e^{\hat i \omega t}.
\label{F_2_i}
\end{equation}
For Eq.~\eqref{F_2_i} to hold,
\begin{equation}
\Psi_{ij}(\omega) = (1 + 2\lambda + \hat i \omega)^{-1}  G_{ij}(\omega).
\label{lambda_sub_freq}
\end{equation}
Additionally, $\sum_\omega \Psi_{ik}(\omega) \Psi_{kj}(-\omega)$ must be a diagonal matrix, which occurs in isotropic flows when the non-identical entries of $\bl \Psi$ are uncorrelated\footnote{While $\Psi_{ik}(\omega) \Psi_{kj}(-\omega)$ for $i\ne j$ is not necessarily smaller than that of $i=j$, $\sum_\omega \Psi_{ik}(\omega) \Psi_{kj}(-\omega)$ for $i\ne j$ will be negligible compared to that of $i=j$ when the sampling period $T$ is sufficiently long as is the case here.}. 
Taking the trace of Eq.~\eqref{F_2_i} yields
\begin{equation}
   \lambda + \lambda^2 + \frac{1}{n} \sum_\omega \Psi_{ij}(\omega) \Psi_{ji}(-\omega) = 0,
\label{lambda_sub}
\end{equation}
where $n$, as defined above, is the number of spatial dimensions.
Due to the long-term incompressibility condition in Eq.~\eqref{continuity}, $G_{ii}(0)$ is zero and does not appear in Eq.~\eqref{lambda_sub}.
From Eqs.~\eqref{lambda_sub_freq} and \eqref{lambda_sub}
\begin{equation}
   \lambda + \lambda^2 + \frac{1}{n} \sum_\omega \frac{G_{ij}(\omega) G_{ji}(-\omega)}{(1 + 2\lambda)^2 + \omega^2}  = 0.
\label{lambda_uns}
\end{equation}
Using the convolution theorem, Eq.~\eqref{lambda_uns} is expressed in terms of a continuous integral as
\begin{equation}
   \lambda + \lambda^2 - \frac{1}{n} \iinf \frac{\tilde \rho^{\rm Q}(\omega)}{(1 + 2\lambda)^2 + \omega^2} \dd \omega = 0,
\label{lambda}
\end{equation}
where $\tilde \rho^{\rm Q}$ is defined in Eq.~\eqref{Q_def}.

\paragraph{Remarks on Eq.~\eqref{lambda}:}
\begin{enumerate}
\item This equation, which will be further developed in Sections \ref{unimodal} and \ref{HIT}, is an integral equation.
Evaluating the integral that appears in this equation requires a knowledge of $\lambda$, which itself is the solution.
Hence, obtaining a closed-form explicit expression for $\lambda$ relies on further simplification and requires additional assumptions.
\item The expressions given by RM and SL are special forms of Eq.~\eqref{lambda}, which can be reproduced exactly by adopting further assumptions. 
Specifically, linearizing Eq.~\eqref{lambda} at the limit of $|\lambda| \ll 1$ and using Eq.~\eqref{C_lambda} to express it in terms of $\mathcal C$ exactly reproduces SL (Eq.~\eqref{SL}).
Furthermore, neglecting $\omega^2$ in the denominator reduces the integral to $\rho^{\rm Q}(0) = \overline{ \|\bl \Omega\|^2 - \|\bl S\|^2}$, exactly reproducing RM (Eq.~\eqref{RM}). 
\item No length scale appears in this equation, justifying our earlier arbitrary choice of $L$ in Section~\ref{intro}.  
The choice of time scale, on the other hand, affects terms with $\omega$, $\lambda$, and $\tilde \rho^{\rm Q}$.
These parameters are all normalized by the particle relaxation time $\tau$ in Eq.~\eqref{lambda}.
In most physical scenarios, $\tilde \rho^{\rm Q}$ is governed primarily by the underlying flow rather than $\tau$, and a natural choice for normalization of Eq.~\eqref{lambda} is $\tau_\eta$.
If normalized based on $\tau_\eta$, St will appear in the denominator of the integral, which is compatible with the notion that the large St particles barely respond to the flow fluctuations.
From that re-normalization, one can show $\lambda_\eta \propto {\rm St}Q_\eta$ for $\rm St \ll 1$ and $\lambda_\eta \propto {\rm St}^{-1/2}|Q_\eta|^{1/4}$ for $\rm St \gg 1$, where subscript $\eta$ denotes normalization based on $\tau_\eta$ (refer to Section~\ref{St-renorm} for more detail).
\item Contrary to RM and SL, $\lambda$ or $\mathcal C$ are nonlinear functions of $\tilde \rho^{\rm Q}$. 
Scaling $\tilde \rho^{\rm Q}$ by a factor of $k$, clouds contract or expand $k$ times faster only if $k \to 0$. 
In the limit $k \to \infty$ and if $\lambda \ne -1/2$, $\lambda$ will scale as $k^{1/4}$. 
The numerical results of Section \ref{numerical} confirm this asymptotic prediction. 
\item Similar to RM and SL, the only term that appears in this equation which relates $\lambda$ to the underlying flow is $\tilde \rho^{\rm Q}$, which is closely related to the second invariant of the velocity gradient tensor. 
As expected, this equation is invariant under Galilean transformation (translation and rotation). 
The absence of other flow-related parameters suggests that particle clustering, although complex, solely depends on the difference between the norms of the autocorrelation function of the rotation- and strain-rate tensors. 
\end{enumerate}

To better understand the implication of Eq.~\eqref{lambda}, it is simplified in the next section to find an explicit expression for $\lambda$ for a case in which the underlying flow oscillates at a single frequency.

\section{One-dimensional unimodal excitation} \label{unimodal}
Earlier, we argued that the clustering phenomenon is complex. 
It exhibits a multiscale behavior (an aspect which is not discussed here) and follows a non-monotonic trend versus Stokes number. 
One may associate this complexity in its entirety to the complexity of the background flow. 
Such an assessment might stem from the fact that turbulent flows, which are often employed for studying clustering, are also very complex. 
On the other hand, one may hypothesize that this complexity is partially a manifestation of the fundamental response of Eq.~\eqref{stokes} to what could be considered a far simpler flow than turbulence.
To distinguish the role of the two and discern the mechanism by which some complexities in the clustering phenomenon arises, we will consider a simple model problem in this section.
Later on, we will show that this simple model problem is fundamentally important in the study of particle clustering as it explains much of our observation when we consider a three-dimensional turbulent flow.

Consider a two-dimensional pure straining flow that oscillates in time with a given frequency and amplitude. That is 
\begin{equation}
\left[ \begin{array}{c} 
u_1 \\ u_2 
\end{array} \right] := \left[ \begin{array}{c} 
x_1 \\ -x_2 
\end{array} \right] \sqrt{-2\Phi} \cos(\omega t),
\label{straining_flow}
\end{equation}
where $\Phi$ and $\omega$ are the root square of amplitude and frequency of the oscillations, respectively, and are independent parameters that are defined based on the underlying flow. 
Suppose $\Phi < 0$ for now, as $\Phi > 0$ is a more complex case to be considered later. 
The factor of $-2$ and square are incorporated into this definition to simplify the notation of the following expressions and allow for a one-to-one correspondence of $\Phi$ and $\tilde \rho^{\rm Q}$. 
For this flow, the velocity gradient tensor is diagonal with $|\partial u_1/\partial x_1| = |\partial u_2/\partial x_2|$. 
The particle motion in the $x$-direction is independent of $u_2$ as is the particle motion in the $y$-direction is independent of $u_1$. 
Owing to the diagonal structure of $\nabla \bl u$ and the equivalence of the rate of strain in $x$- and $y$-directions, one can analyze the motion of a particle in one direction and apply the results to both directions. 
Thus, instead of analyzing two identical problems, we analyze a one-dimensional flow with
\begin{equation}
\frac{\partial u}{\partial x} := \sqrt{- 2\Phi} \cos(\omega t),
\label{1Du_def}
\end{equation}
which, in other words, corresponds to the flow around particles located on the $x$-axis in the two-dimensional setting described above. 

Both straining and rotating flow can be represented in this one-dimensional setting using $\Phi < 0$ and $\Phi > 0$, respectively. 
Interpreting Eq. \eqref{1Du_def} is simple in the straining regime, since the underlying $u(x,t)$ is real and varies linearly along the line with all fluid particles moving away or toward a single point. 
Its interpretation in the rotation regime, on the other hand, requires imagining the line to have an oscillatory rotational motion with all fluid particles oscillating along concentric arches. 
From Eq. \eqref{F_eq}, one can prove that the rotating regime described in Eq. \eqref{1Du_def} with $\Phi > 0$ produces a Lyapunov exponent that is identical to that of a forced vortex defined by $\bl u = [x_2, -x_1] \sqrt{2\Phi} \cos(\omega t)$ in two dimensions.
For this two-dimensional rotating flow, the eigenvalues of the velocity gradient tensor are imaginary and correspond to the single eigenvalue of the one-dimensional model problem.
In contrast to the one-dimensional case, however, $\nabla \bl u$ is real in the two-dimensional case, allowing one to carry out the computations in the real plane.

\begin{figure}
\begin{center}
\includegraphics[width=0.6\textwidth]{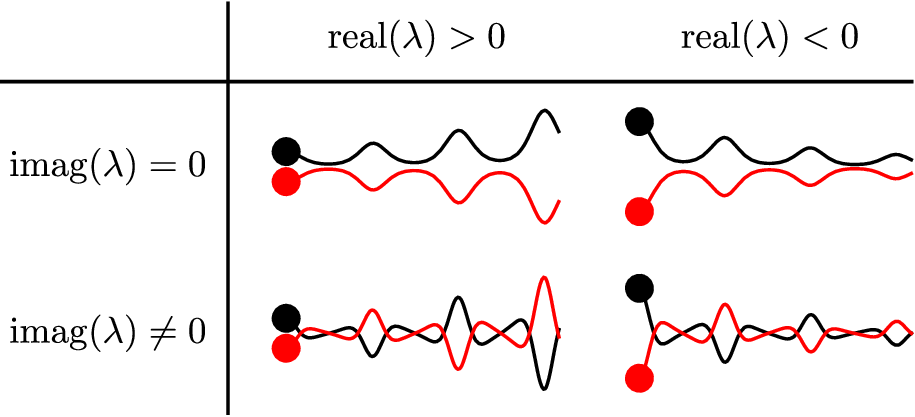}
\caption{Possible relative motion of two particles (black and red) as a function of time in a one-dimensional straining flow.
They can diverge with crossover ($\real\lambda >0$ and $\imag\lambda \ne 0$), diverge without crossover ($\real\lambda >0$ and $\imag\lambda = 0$), converge with crossover ($\real\lambda < 0$ and $\imag\lambda \ne 0$), and converge without crossovers ($\real\lambda < 0$ and $\imag\lambda = 0$).
Similar table can be constructed for a rotating flow (colors online).}
\label{fig:two_prt}
\end{center}
\end{figure}

Pertaining to what was discussed in Section \ref{ana_der}, the long-term rate at which the distance between two particles changes in a cloud is exponential.
To show this in a one-dimensional setting, consider two particles separated by $\delta x$.
The difference between the fluid velocity at the location of these particles is $(\partial u/\partial x)\delta x$.
Since $\partial u/\partial x$ does not scale with $\delta x$, the rate at which particle move relative to each other scales with their distance, leading to an exponential change of $\delta x$.
The exponential rate at which these particles converge or diverge to each other as $t \to \infty$ is by definition the Lyapunov exponent $\lambda$. 
Depending on the value of $\lambda$, four scenarios may occur.
As shown in Figure \ref{fig:two_prt}, particles move relative to each other and may diverge ($\real \lambda > 0$) or converge ($\real \lambda < 0$), representing regimes at which particles contaminate a larger or smaller space over time, respectively.
The convergence or divergence of particles may occur while their trajectories cross ($\imag \lambda \ne 0$) or do not cross ($\imag \lambda = 0$). 
Similar scenarios can be hypothesized for a rotating flow.
Some of these scenarios, however, may never occur in reality (e.g., $\real \lambda < 0$ for heavy particles in a rotating flow). 
Our goal in this section is to find an analytical relationship for $\lambda$ that firstly delineates between the above scenarios and secondly, provides a quantitative estimate of $\lambda$ when Eq.~\eqref{1Du_def} holds.

The one-dimensional flow described above is characterized using Eq.~\eqref{Q_def} and Eq.~\eqref{1Du_def} as
\begin{equation}
\tilde \rho^{\rm Q}(\omega^\prime) = n \Phi \delta(\omega-\omega^\prime),
\label{Phi_def}
\end{equation}
where $\delta$ is the Dirac delta function. 
The sign of $\tilde \rho^{\rm Q}$ reaffirms our earlier distinction between the rotating ($\Phi > 0$) and straining ($\Phi<0$) regimes. 
Substituting $\tilde \rho^{\rm Q}$ into Eq.~\eqref{lambda} and taking $n = 1$ simplifies it to
\begin{equation}
\lambda + \lambda^2 - \frac{\Phi}{(1 + 2\lambda)^2 + \omega^2} = 0.
\label{lambda_4th}
\end{equation}
In total, there are two non-dimensional parameters that appear in Eq.~\eqref{lambda_4th}, which are $\Phi$ and $\omega$.
In Section \ref{intro}, we employed particle relaxation time $\tau$ to normalize all parameters. 
Since $\Phi$ and $\omega$ are normalized by $\tau^2$ and $\tau$, respectively, the effect of Stokes number, viz. $\tau_\eta/\tau$, is embedded in both parameters. 
Equation~\eqref{Phi_def} and consequently Eq.~\eqref{lambda_4th} also represent multi-dimensional isotropic flows as long as $\tilde \rho^{\rm Q}$ contains only a single frequency. 

\begin{figure}
\begin{center}
\includegraphics[width=0.5\textwidth]{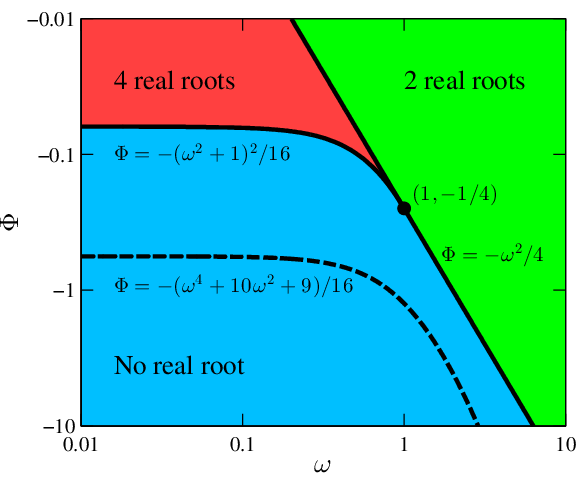}
\caption{The discriminants of Eq.~\eqref{lambda_4th} and its number of real roots for different values of $\Phi$ and $\omega$. 
There are always two real roots for $\Phi > 0$ (not shown) -- (colors online).}
\label{fig:C_NIR}
\end{center}
\end{figure}

Next, we find a relationship for $\lambda = \lambda(\Phi,\omega)$ using Eq.~\eqref{lambda_4th}. 
According to Eq.~\eqref{lambda_4th}, $\lambda$ is one of the roots of a fourth-order polynomial.
Closer examination shows that this polynomial has two real roots for $\Phi > -\omega^2/4$ or $\Phi > 0$, no real roots for $\Phi < -(\omega^2+1)^2/16$ or $-(\omega^2+1)^2/16 < \Phi < -\omega^2/4$, and four real roots otherwise (Figure \ref{fig:C_NIR}).
We show below that out of four roots of this polynomial, only one is physically relevant.

\begin{figure}
\begin{center}
\includegraphics[width=1.0\textwidth]{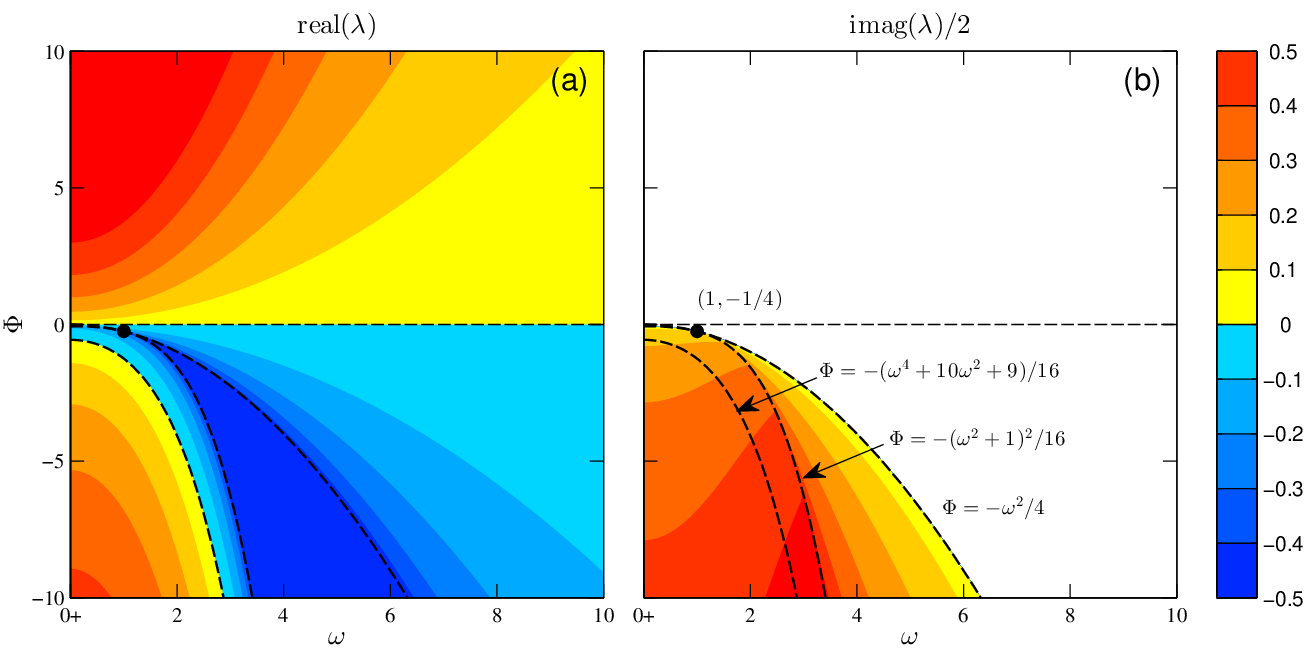}
\caption{The rate of expansion or contraction $\real \lambda$ (a) and crossover $\imag \lambda$ (b) of a pair of particles predicted by Eq.~\eqref{lambda_1D} for a flow that is described by Eq.~\eqref{1Du_def}.
Expansion ($\real\lambda > 0$) is predicted in rotating ($\Phi > 0$) or strong straining ($\Phi \ll 0$) flows.
For $-(\omega^2+1)^2/16 < \Phi < -\omega^2/4$ and $\omega > 1$, $\real\lambda$ is constant $-1/2$.
Particles crossover only in the straining regime when $\Phi < \max \left[-(\omega^2+1)^2/16,\min\left(-\omega^2/4,-1/4\right)\right]$ (colors online).}
\label{fig:C1a}
\end{center}
\end{figure}

The roots of Eq.~\eqref{lambda_4th} can be analytically computed by converting it to a depressed quadratic form and taking 
\begin{equation}
\gamma := 1 + 2\lambda.
\label{gam_def}
\end{equation}
With this change of variable, Eq.~\eqref{lambda_4th} becomes
\begin{equation}
\gamma^4 + \left(\omega^2 - 1\right)\gamma^2 - 4\Phi - \omega^2 = 0.
\label{d_4th}
\end{equation}
Therefore,
\begin{equation}
\gamma^2 = \frac{1}{2}(1-\omega^2) \pm \frac{1}{2}\sqrt{(\omega^2+1)^2 + 16\Phi}.
\label{gam_s}
\end{equation}
No contraction or expansion is expected in the absence of flow.
Thus, the condition $\lambda = 0$, which corresponds to $\gamma = 1$, must be satisfied when $\Phi = 0$.
Therefore, from the two possible solutions in Eq.~\eqref{gam_s}, only the root corresponding to the plus sign is admissible.
Computing $\lambda$ from Eqs.~\eqref{gam_def} and \eqref{gam_s} and imposing the condition $\lambda(\Phi=0,\omega) = 0$ for one more time gives
\begin{equation}
\lambda = -\frac{1}{2} + \frac{\sqrt{2}}{4}\sqrt{1-\omega^2 + \sqrt{(\omega^2+1)^2 + 16\Phi}},
\label{lambda_1D}
\end{equation}
which is valid for $\omega$ and $\Phi \in \mathbb R$.
Based on this equation $\lambda$ can be complex for a specific combination of $\Phi$ and $\omega$, pertaining to the occurrence of particle crossovers.
The magnitude of $\imag \lambda$ is crossover frequency and proportional to the number of times particles exchange side in a unit time. 
The real and imaginary part of $\lambda(\Phi,\omega)$ are plotted in Figure \ref{fig:C1a}.

\paragraph{Remarks on Eq.~\eqref{lambda_1D}:}
\begin{enumerate}
\item $\real\lambda$ represents the rate of expansion ($\real\lambda > 0$) or contraction ($\real\lambda < 0$).
Since the second term in Eq.~\eqref{lambda_1D} is always positive, the strongest possible rate of contraction is $-1/2$.
For $\omega < 1$, the minimum value of $\real\lambda$ occurs at the discriminant curve $\Phi = -(\omega^2+1)^2/16$, at which $\real\lambda$ is not differentiable.
For $\omega > 1$, the minimum occurs in a region enclosed between $-(\omega^2+1)^2/16 < \Phi < -\omega^2/4$, where $\real{\lambda(\Phi,\omega)}$ is constant and equal to the global minimum $-1/2$ (Figure \ref{fig:C1a}).
\item No expansion or contraction is predicted for two cases.
The first is the trivial case in which $\Phi = 0$.
The second case occurs at $\Phi = -(\omega^4 + 10\omega^2 + 9)/16$.
In the latter case only $\real\lambda = 0$ and $\imag\lambda \ne 0$, indicating pure oscillation of the volume of the clouds between positive and negative values (Figure \ref{fig:two_prt}).
\item In the rotation dominated regimes where the flow is elliptical and $\Phi > 0$, the clouds of particles always experience pure expansion with $\real\lambda > 0$.
No crossovers occurs in this regime (Figure \ref{fig:C1a}).
\item Contrary to the contraction-rate, the expansion-rate is unbounded.
In general $\real\lambda$ is proportional to $\Phi$ for $|\Phi| \ll 1$ and to $|\Phi|^{1/4}$ for $|\Phi| \gg 1$.
\item As discussed above, $\imag\lambda\ne0$ is the byproduct of $\det{\bl J} < 0$, occurring in strong-straining flows as the cloud fully collapses and turns inside-out.
In the one-dimensional setting, $\imag\lambda \ne 0$ occurs when two particles exchange sides on the line.
\item Particle crossover occurs when $\Phi < \max \left[-(\omega^2+1)^2/16,\min\left(-\omega^2/4,-1/4\right)\right]$ (blue region in Figure \ref{fig:C_NIR}).
The maximum of $\imag\lambda$ occurs at $\Phi = -(\omega^2+1)^2/16$ for $\omega > 1$, at which $\imag\lambda$ is not differentiable.
\item Out of four scenarios in Figure \ref{fig:two_prt}, only three are predicted to occur in the one-dimensional straining flow and one in the one-dimensional rotating flow.
In a straining regime $\{\real \lambda > 0, \; \imag \lambda = 0\}$  never occurs while in a rotating regime only $\{\real\lambda > 0,\; \imag \lambda = 0\}$ occurs.
In other words, particles have to cross each other to diverge in a straining flow, whereas they may or may not cross when they converge.
Thus, in a straining flow, particle dispersion only occurs if particle trajectories cross, whereas they can cluster regardless of the occurrence of crossovers.
In a rotating flow, they always diverge without crossing over each other.
\item For a straining flow oscillating at sufficiently small $\omega$, the onset of crossover occurs at $\Phi=-1/16$ whereas the onset of expansion occurs at $\Phi=-9/16$.
   Therefore, particles that disperse in a straining flow are at least several times more inertial than those which do not cross each other.
\end{enumerate}

Equation \eqref{lambda_1D} is the exact solution of Eq.~\eqref{lambda} when the flow is one-dimensional and excited at a single frequency.
Thus, the underlying assumption of Eq.~\eqref{lambda_1D} is that of Eq.~\eqref{lambda}.
Namely, the high-order oscillatory terms, which originally appeared in Eq.~\eqref{F_1_i}, are neglected by assuming $\|\bl \Psi\|\ll 1$.
To evaluate the significance of this assumption, we compare the prediction of Eq.~\eqref{lambda_1D} to the numerical results in Section~\ref{numerical}.
Before that, however, we discuss the interpretation of these findings when they are expressed as a function of the Stokes number.

\subsection{St-dependent re-normalization of Eqs.~\eqref{lambda} and \eqref{lambda_1D}} \label{St-renorm}
All the derived equations so far have been normalized based on the particle relaxation time $\tau$.
The choice of $\tau$ for the time scale led to compact equations in which only an oscillation amplitude and frequency appear.
This normalization translates to experiments in which a single class of particles is reused in a variety of flows. 
Often in practice, however, we encounter multiple classes of particles in one particular flow. 
To predict the trends observed in the latter case, we normalize the previous results by a flow-dependent time scale rather than $\tau$. 
In this section, by changing the normalization parameter, we show the effect of St on $\mathcal C$ in a one-dimensional pure straining and rotating flows as well as a general arbitrary multi-dimensional flow. 

The Lyapunov exponent $\lambda$ is related to the underlying flow in Eq.~\eqref{lambda} through $\tilde \rho^{\rm Q}$.
Its dimensional counterpart, $\tilde \rho^{\rm Q_{\rm d}}$ is computed based on $\overline{\|\bl S_{\rm d}\|}$ and $\overline{\|\bl \Omega_{\rm d}\|}$, where subscript d denotes dimensional variables.
The Kolmogorov time scale $\tau_\eta = \overline{\|\bl S_{\rm d}\|}^{-1}$ thus emerges as the most natural choice for normalization of $\tilde \rho^{\rm Q_{\rm d}}$. 
Employing subscript $\eta$ to distinguish parameters that are normalized based on $\tau_\eta$, we have $\tilde \rho^{\rm Q} = {\rm St} \tilde \rho^{\rm Q_\eta}$, $\lambda = {\rm St} \lambda_\eta$, and $\mathcal C = {\rm St} \mathcal C_\eta$ and $\Phi = {\rm St}^2 \Phi_\eta$.
Thus, re-nondimensionalization of Eq.~\eqref{lambda_1D} based on $\tau_\eta$ yields 
\begin{equation}
\lambda_\eta = -\frac{1}{2\rm {St}} + \frac{\sqrt{2}}{4}\sqrt{{\rm St}^{-2} - \omega_\eta^2 + \sqrt{({\rm St}^{-2} + \omega_\eta^2)^2 + 16{\rm St}^{-2}\Phi_\eta}}.
\label{lambda_1D_St}
\end{equation}

\begin{figure}
\begin{center}
\includegraphics[width=0.6\textwidth]{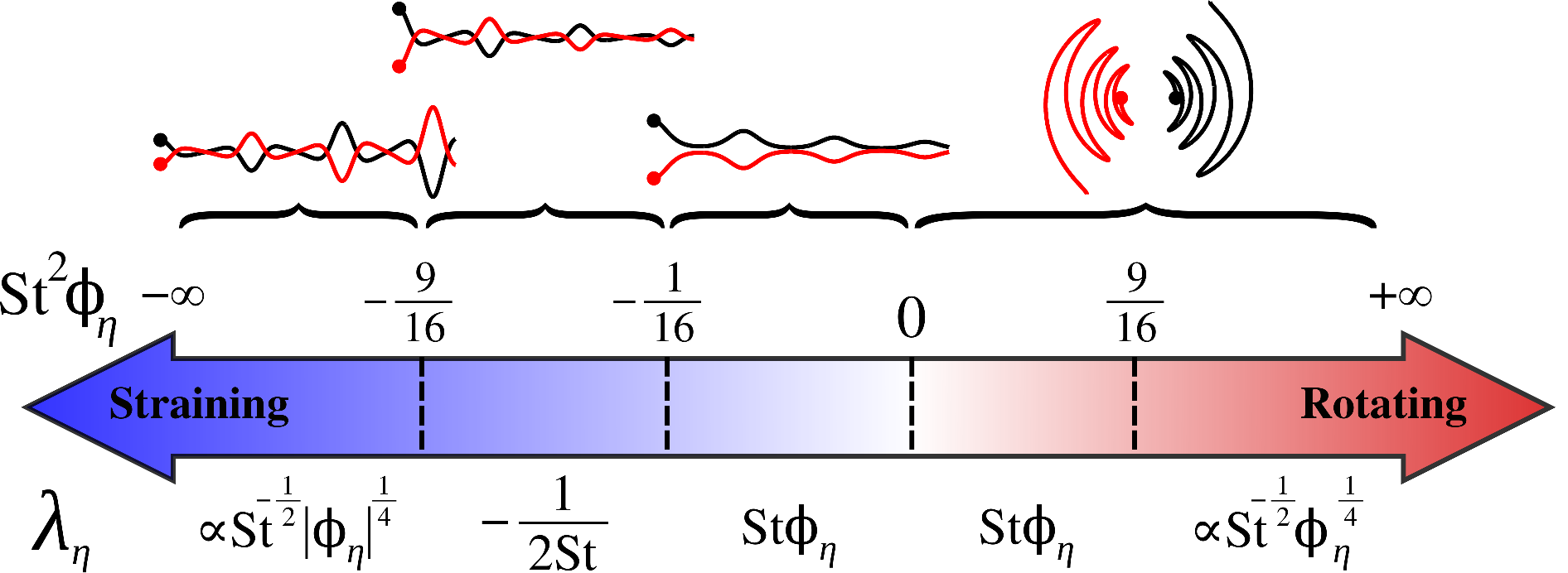}
\caption{The overall behavior of $\lambda_\eta$ (the Lyapunov exponent normalized by the flow time scale) for particles with St$=\tau/\tau_\eta$ in a one-dimensional flow described by Eq.~\eqref{lambda_1D_St} when $\omega_\eta \approx 0$.
Several regimes of $\lambda_\eta$ are distinguished under the arrow, showing where clustering and dispersion occurs.
The transition between regimes is marked by the dashed-line and occurs at the corresponding values of ${\rm St}^2 \Phi_\eta$ shown on the top.
At finite $\omega_\eta$, $-9/16$ and $-1/16$ transitional points approximately translate to $-((2\omega_\eta{\rm St})^2+3)^2/16$ and $-(\max((2\omega_\eta{\rm St})^2,1)/16$, respectively. 
The schematics on the top show how two nearby particles move relative to each other in each regime.
This schematic can be generalized to an arbitrary flow by taking $\Phi_\eta = Q_\eta/n$.}
\label{fig:p-space}
\end{center}
\end{figure}

$\lambda_\eta$, as oppose to $\lambda$, is often reported in the literature when analyzing the clustering of inertial particles (e.g., ~\cite{bec2006lyapunov}).
Similar to the remarks at the end of the last section, one can analyze the behavior of $\lambda_\eta = \lambda_\eta({\rm St}, \Phi_\eta, \omega_\eta)$ under various circumstances.
For the sake of brevity, we have condensed this information in a schematic (Figure \ref{fig:p-space}) for cases in which $\omega_\eta \approx 0$ and postpone its discussion to Section~\ref{1D_SR}, where we compare the results of our one-dimensional model problem to that of a turbulent flow.

The asymptotic predictions in Figure~\ref{fig:p-space}, which are obtained from Eq.~\eqref{lambda_1D_St} for a one-dimensional unimodal excitation case, can be generalized to an arbitrary multi-dimensional flow by taking $\Phi_\eta = Q_\eta/n$.
To show this, consider Eq.~\eqref{lambda} that holds for the general case.
After re-normalization based the Kolmogorov time scale $\tau_\eta$
\begin{equation}
   \frac{\lambda_\eta}{\rm St} + \lambda_\eta^2 - \frac{1}{n} \iinf \frac{\tilde \rho^{\rm Q_\eta}}{(1 + 2\lambda_\eta {\rm St})^2 + (\omega_\eta {\rm St})^2} \dd \omega_\eta = 0.
\label{lambda_St}
\end{equation}
To obtain the asymptotic behavior of $\lambda_\eta$, suppose $\lambda_\eta \propto {\rm St}^p$ where $p$ is an exponent to be determined. 
Assuming $p>-1$ for ${\rm St} \ll 1$, then the second term in Eq.~\eqref{lambda_St} can be neglected compared to the first term, and 1 will be the leading order term in the denominator of the integrant.
Thus, the integral reduces to $Q_\eta$, which implies $\lambda_\eta = {\rm St}Q_\eta/n$ and $p=1$.
The analysis for ${\rm St} \gg 1$ is similar. 
Again assuming $p>-1$, this time the first term can be neglected compared to the second term. 
Assuming the energetic frequencies of $\tilde \rho^{\rm Q_\eta}$ are at low frequencies\footnote{This assumption, which is in agreement with the results obtained from a turbulent flow in Figure~\ref{fig:rQt3}-b, implies very heavy inertial particles only respond to the largest structures of the flow since $\omega \sim \dot x/L$ with $\dot x$ and $L$ being the typical velocity of particle and the size of a flow structure that significantly influences the particle motion, respectively.}, then the second term in the denominator of the integrant can be neglected.
Thus, the integral scales as $({\rm St} \lambda_\eta)^{-2}Q_\eta/n$, implying $\lambda_\eta \propto {\rm St}^{-1/2} |Q_\eta/n|^{1/4}$ and $p=-1/2$.
These results, which are consistent with those of the one-dimensional analysis in Figure~\ref{fig:p-space}, are also confirmed by our numerical results in Section~\ref{HIT}.

\subsection{Numerical validation} \label{numerical}
To validate the present analysis and compute $\lambda$, we ideally need to compute $\bl J(t)$ as $t\to \infty$ provided that $\mathcal C$ (and thus $\lambda$) is defined by Eq.~\eqref{C_def}. 
In practice, however, $\det{\bl J}$ grows exponentially in time, producing an ill-conditioned system for a long integration period. 
To overcome this shortcoming, we infer the long-term response of $\bl J$ to excitation $\partial \bl u/\partial \bl x$ from a single cycle of excitation, as outlined below. 

For either the one-dimensional unimodal case discussed in Section~\ref{unimodal} or the three-dimensional turbulence that will be discussed in Section~\ref{HIT}, $\partial \bl u/\partial \bl x$ will be a harmonic function with period $T$. 
Evidently, $T=2\pi/\omega$ for the one-dimensional flow. 
For the turbulent flow, which is chaotic, $T\to\infty$. 
However, our numerical result shows that if we take $T\gg 1$ for the turbulent flow, i.e., when the dimensional sampling period is much larger than particle relaxation time, the results become independent of $T$. 
Now, if we denote $\bl J(t=NT)$ by $\bl J^{(N)}$, our goal is to compute $\det{\bl J^{(N)}}$ as the number of cycles $N\to\infty$. 
A brute force approach of integrating Eq.~\eqref{J_stokes} for many cycles will produce a large numerical error. 
A better alternative is to compute a transformation matrix $\bl A \in \mathbb R^{2n\times 2n}$ that satisfies
\begin{equation}
\left[\begin{matrix}\bl J^{(1)} \\ \bl {\dot J}^{(1)}\end{matrix}\right] = \bl A \left[\begin{matrix}\bl J^{(0)} \\ \bl {\dot J}^{(0)}\end{matrix}\right].
\label{A_trans}
\end{equation}
Then, since $\det{\bl J^{(0)}} = 1$, $\det{\bl J^{(N)}}$ can be computed as $(\lambda_1^A\lambda_2^A\cdots\lambda_n^A)^N$, where $\lambda_1^A$, $\lambda_2^A$, $\cdots$, $\lambda_n^A$ are the largest $n$ eigenvalues of $\bl A$.
The remaining eigenvalues of $\bl A$ are associated with the decay of the initial conditions and are thus excluded from our calculations. 
Thus, $\mathcal C$, which is $\ln\left(\det{\bl J^{(N)}}\right)/(NT)$ as $N\to\infty$, can be computed as $\ln(\lambda_1^A\lambda_2^A\cdots\lambda_n^A)/T$.

In practice, we compute $\bl A$ column-by-column by selecting $2n$ linearly independent initial conditions for $[\begin{matrix}\bl J & \bl {\dot J}\end{matrix}]^{\rm T}$ and calculating its time evolution for $T$ using Eq.~\eqref{J_stokes}.
$\partial \bl u/\partial \bl x$ in Eq.~\eqref{J_stokes} is prescribed based on Eq.~\eqref{1Du_def} for the unimodal excitation case.
As discussed in detail in Section~\ref{HIT}, it is numerically extracted from the direct numerical simulation for the three-dimensional turbulence by recording its temporal variation along particle trajectories. 
All calculations are performed in the complex plane using the fourth-order Runge-Kutta time integration scheme.
The time step size, which has been verified to be sufficiently small, is $2\pi \times 10^{-6}/\omega$ for the one-dimensional and a fraction of $\tau_\eta$ for the three-dimensional calculations. 
We refer to $\lambda$ (or $\mathcal C$) obtained from this procedure as the reference numerical result and use it to validate our analysis.

\begin{figure}
\begin{center}
\includegraphics[width=0.5\textwidth]{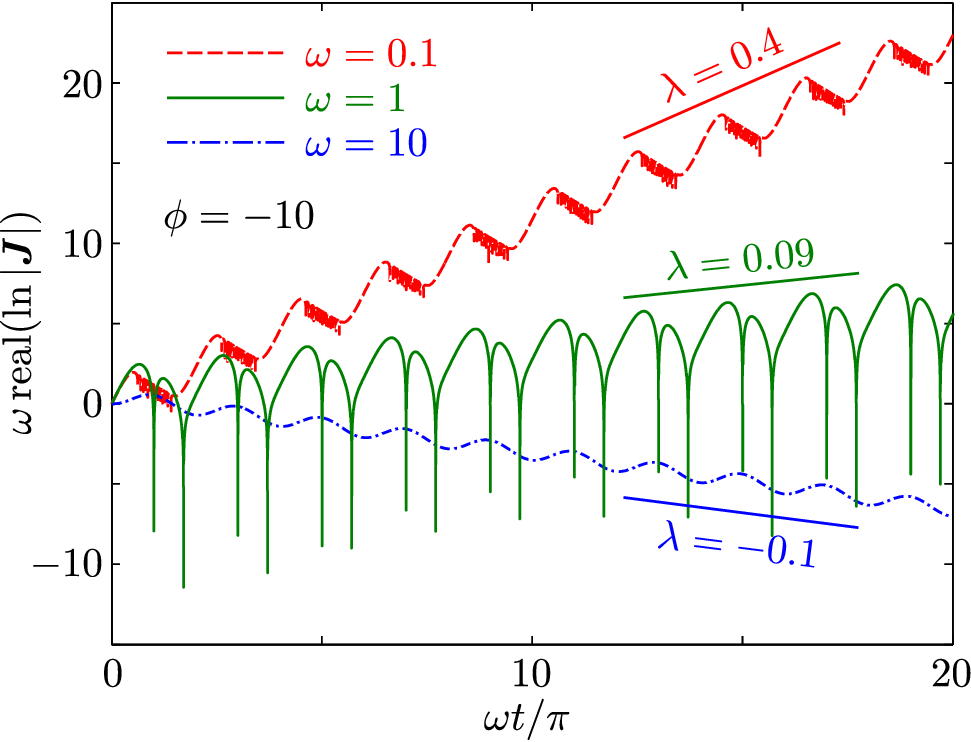}
\caption{The time variation of the distance between two particles $\det{\bl J}$ subjected to an oscillatory velocity gradient (Eq. \eqref{1Du_def}) at three frequencies. 
Curves are obtained from the numerical integration of Eq.~\eqref{J_stokes}.
The time-averaged slope of these curves provides a numerical estimate for $\lambda$.
The spikes in these curves correspond to $\det{\bl J}=0$ associated with particle trajectory crossing (colors online).}
\label{fig:y_vs_t}
\end{center}
\end{figure}

For the one-dimensional case, the behavior of $\ln (\det{\bl J})$ as a function of time varies depending on $\omega$ and $\Phi$ significantly (Figure \ref{fig:y_vs_t}). 
Based on Eq.~\eqref{C_cal}, the long-term slope of these curves is equivalent to $\lambda$ (or $\mathcal C$ since $n=1$).
These one-dimensional calculations are also repeated for $n = 2$ with non-diagonal $\nabla \bl u$ representing a forced vortex.
The results are identical to that of the one-dimensional rotating flow, showing the applicability of Eq.~\eqref{Phi_def} to the higher-dimensional unimodal isotropic flows.
Spikes in Figure \ref{fig:y_vs_t} corresponds to the occurrence of particle crossover when the distance between two particles and $\det{\bl J}$ becomes zero and changes sign. 
The occurrence of these incidences can be readily deduced from the homogeneous form of Eq.~\eqref{F_eq}.
When $F < -1$, then $\det F = -F^2-F < 0$, leading to unstable growth of $F$ to $-\infty$ that indicates $\det{J} = 0$. 
These crossover incidents, which has been called the sling effect in the literature, have been predicted theoretically and shown experimentally in the past \cite{falkovich2002acceleration, bewley2013observation}.

\begin{figure}
\begin{center}
\includegraphics[width=1.0\textwidth]{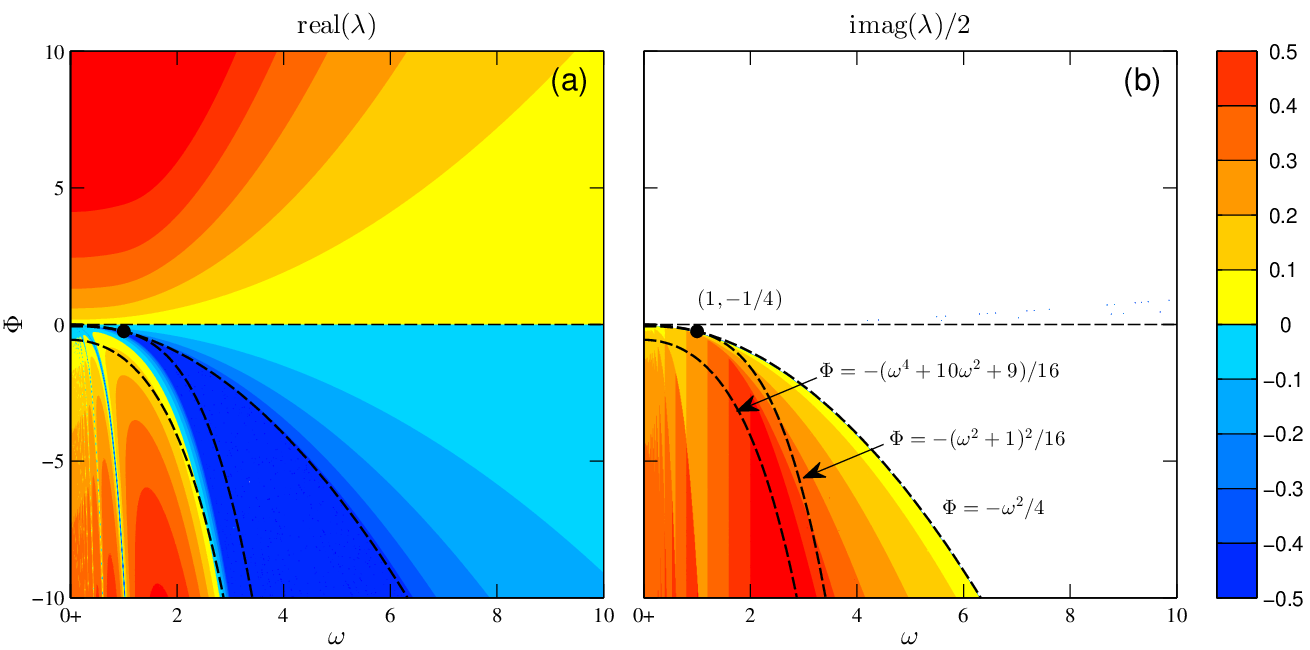}
\caption{The rate of expansion or contraction $\real \lambda$ (a) and crossover $\imag \lambda$ (b) of a pair of particles for a flow oscillating at frequency $\omega$ with a strain- or rotation-rate $\Phi$. 
   $\Phi>0$ and $\Phi<0$ represent rotating and straining flows, respectively. 
   These results are obtained from the numerical integration of Eq.~\eqref{F_eq}.
   Dashed lines are predicted discriminants from our analysis and replicated from Figure \ref{fig:C1a} (colors online).}
\label{fig:C1n}
\end{center}
\end{figure}

Following the above procedure, $\lambda$ is computed for $\omega \in (0,10]$ and $\Phi \in [-10,10]$ on a $1024\times1024$ discrete parameter space.
These numerical calculations show that $\lambda$ is a highly nonlinear function of $\omega$ and $\Phi$ (Figure \ref{fig:C1n}).
For a certain combination of $\omega$ and $\Phi$, $\lambda$ is not differentiable or has a very sharp gradient.
These results confirm the possibility of expansion (dispersion) in straining flows, where $\Phi < 0$ and $\real \lambda > 0$.
They also confirm the existence of a bound on the contraction-rate that never drops below $-1/2$.
The asymptotic behaviors $\lambda \propto |\Phi|^{1/4}$ as $|\Phi| \to \infty$ and $\lambda \propto \Phi$ when $\Phi \ll 1$ are also confirmed by the numerical results (Figure \ref{fig:p-space}).
These numerical observations are in full agreement with the prediction of our analysis (Figures \ref{fig:C1a} and \ref{fig:C1n}).

The numerically computed $\imag\lambda$ is also nonzero for a certain combination of $\omega$ and $\Phi$.
In those cases, $\imag\lambda$ can be either positive or negative depending on the initial condition. 
Its magnitude, however, is unique and depends on the crossover frequency. 
According to Figures \ref{fig:C1a} and \ref{fig:C1n}, there is an agreement between the numerical and analytical results in terms of $\omega$ and $\Phi$ at which $\imag\lambda \ne 0$, viz. the blue region in Figure \ref{fig:C_NIR}.
This agreement shows the ability of our analysis in predicting the regimes of particle crossover correctly. 

\begin{figure}
\begin{center}
\includegraphics[width=1.0\textwidth]{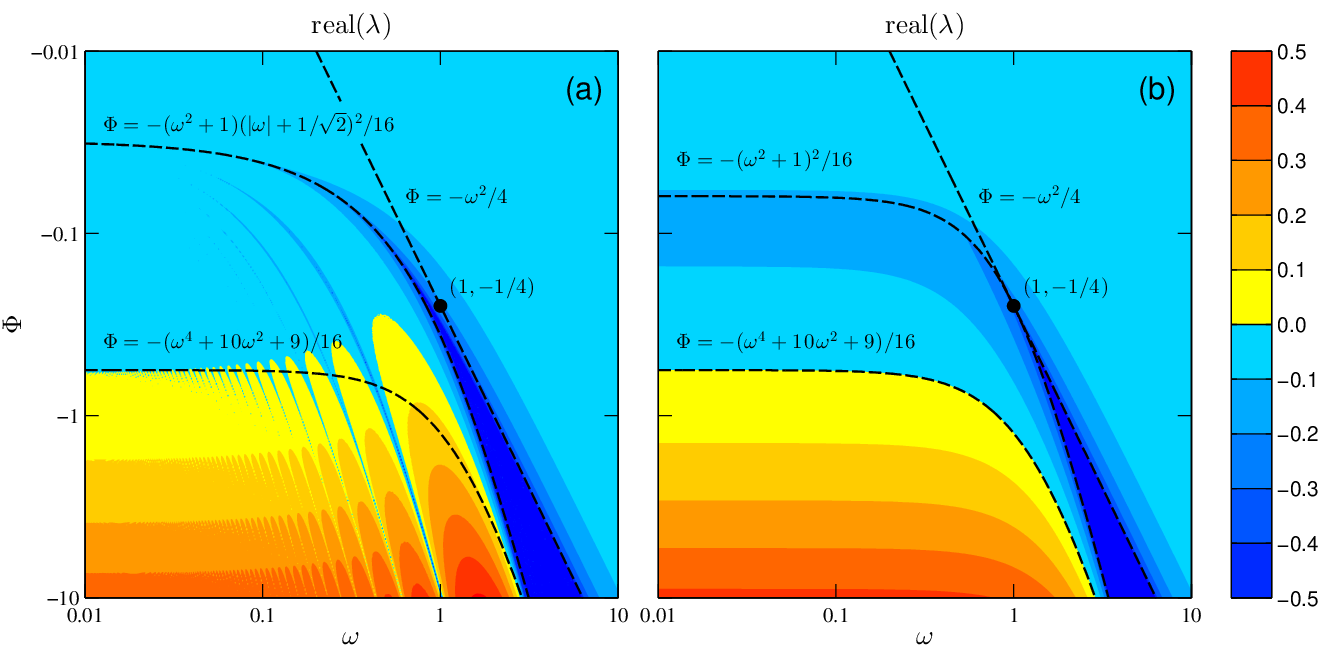}
\caption{The rate of expansion or contraction $\real\lambda$ in a straining regime with $\Phi < 0$, obtained from the numerical result (a) and the present analysis (b) in a logarithmic scale. 
All the dashed lines are extracted from our analysis with the exception of $\Phi = -(\omega^2+1)(|\omega| + 1/\sqrt{2})^2/16$, which is obtained via curve-fitting to the numerical result (colors online).}
\label{fig:C1_log}
\end{center}
\end{figure}

Discrepancies between the numerical and analytical results are also observed, specifically for $\Phi \lesssim -1$ and $\omega \lesssim 1$ where $\imag\lambda \ne 0$ (blue region in Figure \ref{fig:C_NIR}). 
A higher degree of non-linearity is observed in the numerical result for $\Phi < -(\omega^4+10\omega^2+9)/16$. 
Additionally for $\omega > 1$, while our analysis predicts $\real\lambda = -1/2$ for $-(\omega^2+1)^2/16 < \Phi < -\omega^2/4$, the numerical result shows a broader range of $\Phi$ producing $\real\lambda = -1/2$. 
Fitting a curve to the numerical result shows $\real\lambda = -1/2$ for $-(\omega^2+1)(|\omega| + 1/\sqrt{2})^2/16 < \Phi < -\omega^2/4$, which includes $\Phi$ slightly lower than the analytical prediction (Figures \ref{fig:C1_log}).
In overall, however, the discriminant curves obtained from our analysis provide a good approximation for the values at which the numerical solution changes sign or reaches a plateau. 

$\Phi$ and $\omega$ are normalized by the particle relaxation time $\tau$ and thus, are proportional to St. 
This correspondence implies that $\Phi\gg1$ and $\omega\gg1$ represent particles with a high St. 
As a result, much of the depicted parameter space in Figures \ref{fig:C1a} and \ref{fig:C1n} is relevant to high St regime.
Our analysis captures the general trend of $\lambda(\Phi,\omega)$ in this part of the parameter space.
However, extremely narrow valleys of parameter space with $\real\lambda < 0$ (e.g. $\Phi = -10$ and $\omega = 1$) are missing in the analytical result (Figure \ref{fig:C1_log}).
The pattern produced by these valleys resembles a fractal structure. 
The distance between valleys reduces, and they become shallower as $\omega \to 0$.
For a given $\omega < 1$, these valleys appear only at $\Phi < -(\omega^2+1)(|\omega| + 1/\sqrt{2})^2/16$ (Figure \ref{fig:C1_log}). 
These valleys may override the extremum of $\real\lambda$ predicted at $\Phi = -(\omega^2 + 1)^2/16$ by our analysis.
As discussed in detail in Section \ref{1D_SR}, the predicted extremum agrees with the numerical result when $\omega \to 0$. 
Neglecting the local extremum associated with the valleys at finite $\omega < 1$, the remainder of $\real{\lambda(\omega,\Phi)}$ is a smooth envelope that has an extremum at the predicted $\Phi = -(\omega^2 + 1)^2/16$.

The discovery of these narrow valleys has an important implication in designing new hydrodynamic particle separators with high precision. 
One can generate a pure oscillatory straining flow laden with particles of various density or size and adjust the oscillation frequency such that some particles cluster and others disperse in time. 
Note that $\omega$ and $\Phi$ in Figures \ref{fig:C1n} and \ref{fig:C1_log} are normalized by $\tau$ and $\tau^2$, respectively. 
Hence, the dispersion or clustering of particles with different $\tau$ in a given oscillatory pure straining flow is determined based on the value of $\real \lambda$ extracted from a parabolic curve in Figures \ref{fig:C1n} and \ref{fig:C1_log}. 
This parabolic curve can intersect with the valleys, which occur in a part of parameter space where $\Phi$ is a quartic function of $\omega$. 
By adjusting the oscillation frequency of the flow, the intersection point can be changed such that $(\omega,\Phi)$ of a pre-specified particle class coincides with a valley. 
As a result, while that pre-specified particle class clusters over time, particles with slightly higher or lower $\tau$ will disperse as their $(\omega,\Phi)$ is outside of the valley. 
To demonstrate this idea, we simulated a 2D pure oscillatory straining flow laden with two class of particles.
Particles in these two classes are different in size merely by 1\%.
The flow parameters are adjusted according to the above procedure to have $(\omega,\Phi)$ of particles inside and outside of the valley that passes through $(\omega,\Phi)=(1,-10)$. 
An animation in the supplementary material shows the response of these two classes of particles (red and black) to the flow.
Only after a few oscillations, one class of particles clusters and accumulates toward the flow stagnation point whereas the other class disperses and gets removed from the oscillatory region of the flow. 
In practice, neither the drag law or the flow field is perfectly linear. 
The particles have a finite size and density and may interact with each other or the solid boundaries of the apparatus. 
Although our preliminary numerical results show that the existence of these narrow valleys is robust against these imperfections, an experimental study is still needed to confirm the feasibility of this novel concept for particle separation in the future.

\subsection{Comparison against other models} \label{comp}

The present analysis is a more general form of our former \cite{esmaily2016CPT} and Maxey's \cite{maxey1987gravitational} analysis.
For the case of one-dimensional unimodal excitation, SL is simplified by combining Eqs.~\eqref{SL}, \eqref{C_lambda}, and \eqref{Phi_def} as
\begin{equation}
\lambda = \frac{\Phi}{1 + \omega^2},
\label{SL_1D}
\end{equation}
and similarly, RM is simplified by combining Eqs.~\eqref{RM}, \eqref{C_lambda}, and \eqref{Phi_def} as
\begin{equation}
\lambda = \Phi. 
\label{RM_1D}
\end{equation}
These two relationships can be derived from Eq.~\eqref{lambda_1D} or Eq.~\eqref{d_4th} under the assumptions of $|\lambda|\ll 1$ and $|\omega|\ll 1$ as well, therefore they can be considered a special case of the present analysis.

\begin{figure}
\begin{center}
\includegraphics[width=1.0\textwidth]{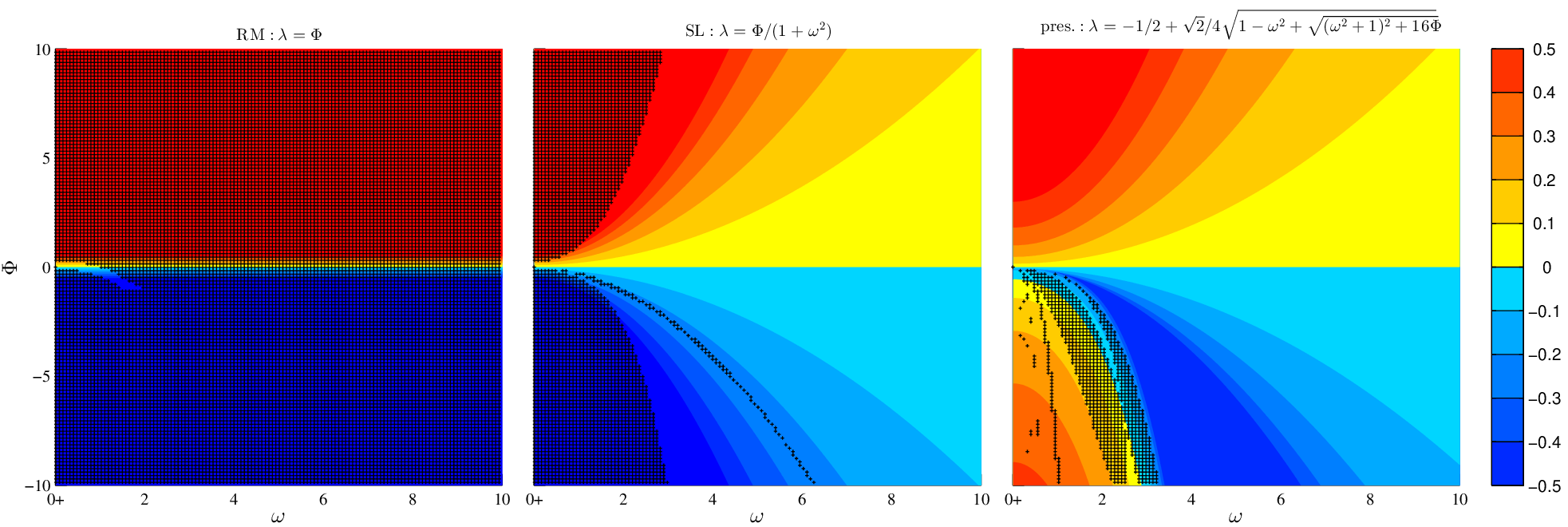}
\caption{The rate of expansion or contraction $\real\lambda$ predicted from RM (left), SL (center), and the present analysis (right), as a function of $\omega$ and $\Phi$.
The hatched area denotes parts of the parameter space where the predicted $\real\lambda$ is larger or smaller than that of the reference by a factor of 2.
Note, RM is only valid near the origin (colors online).}
\label{fig:valid_lambda}
\end{center}
\end{figure}

\begin{figure}
\begin{center}
\includegraphics[width=1.0\textwidth]{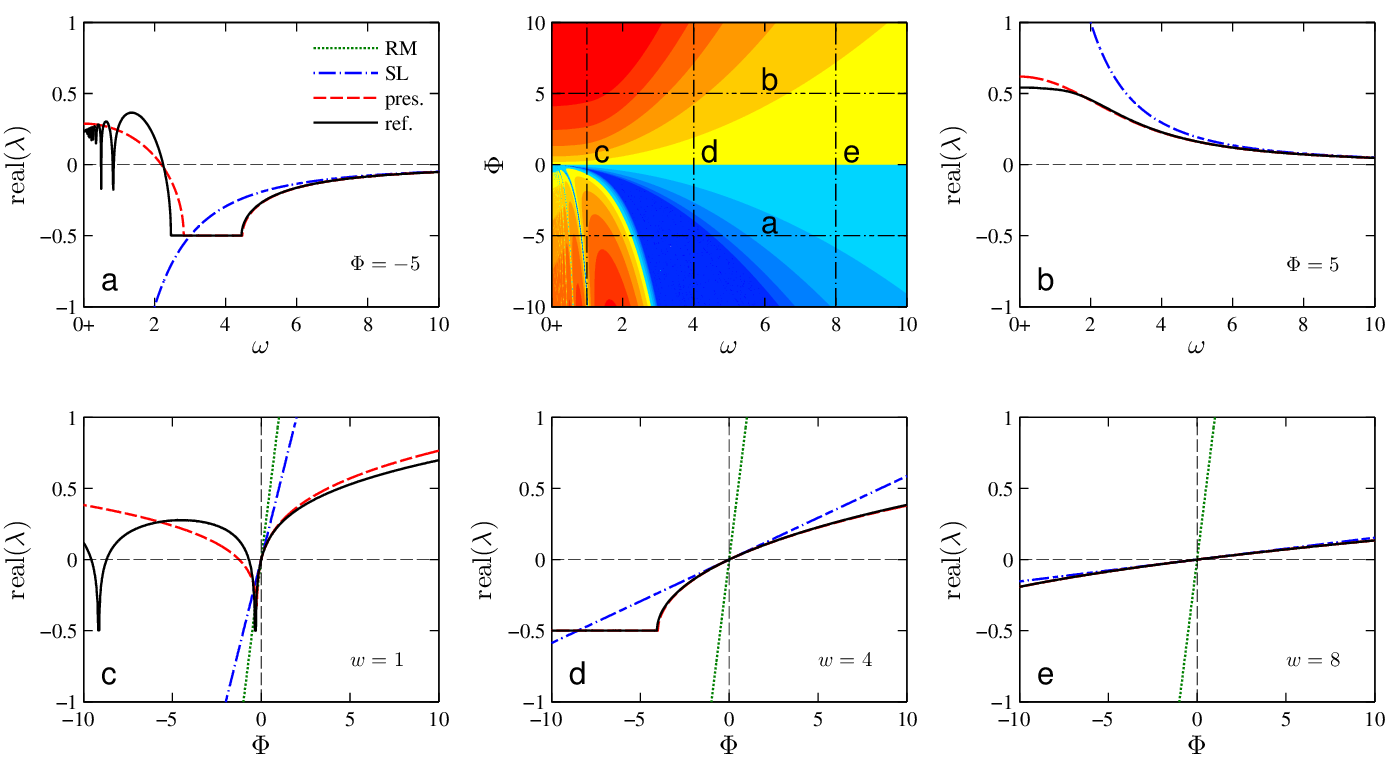}
\caption{The rate of expansion or contraction $\real\lambda$ from the numerical calculations (solid black), the present analysis or Eq.~\eqref{lambda_1D} (dashed red), SL or Eq.~\eqref{SL_1D} (dashed-dotted blue), and RM or Eq.~\eqref{RM_1D} (dotted green), at different values of $\omega$ and $\Phi$.
The prediction of RM is outside of the depicted range in (a) and (b) and not shown.
The prediction of the present analysis is not visible in some plots as it fully collapses with the reference result (colors online).}
\label{fig:C_comp}
\end{center}
\end{figure}

RM and SL are valid only within the regimes compatible with their underlying assumptions (Figure \ref{fig:valid_lambda}).
RM is derived for small St, which translates to $\omega \ll 1$ and $|\Phi| \ll 1$ (unless $|\Phi| \approx 0$, where there is no flow and $\lambda \approx 0$).
SL is derived for small $\real \lambda$, which translates to $\omega \gg 1$ or $|\Phi| \ll 1$.
The present analysis, on the other hand, provides a reasonable approximation at all $\omega$ and $\Phi$.
To highlight these differences, we have compared the predictions of all models against the reference numerical results in Figures \ref{fig:valid_lambda} and \ref{fig:C_comp}.
Figure \ref{fig:C_comp}(c) confirms RM and SL predictions are invalid at small $\omega$, except for $\Phi \ll 1$. 
For example, these two relations predict $\lambda=-5$ at $\omega\approx 0$ and $\Phi=-5$, whereas $\lambda=0.3$ and 0.29 from the numerical calculations and our analysis, respectively.
At higher $\omega$, SL prediction approaches the numerical solution for a wide range of $|\Phi|$, while RM prediction is wrong everywhere except $|\Phi| \approx 0$ (Figure \ref{fig:C_comp}(e) and Figure \ref{fig:valid_lambda}).

The nonlinear behavior of $\real \lambda$ at high St regime is captured only by our analysis. 
According to Eqs. \eqref{SL_1D} and \eqref{RM_1D}, RM and SL are both linear functions of $\Phi$.
Hence, they both fail to predict the nonlinear behavior of $\real\lambda$ at higher $|\Phi|/\omega$ (Figures \ref{fig:C1n} and \ref{fig:C_comp}).
Neither RM or SL predicts the lower bound on the contraction rate ($\min(\lambda)\ge -0.5$), possibility of expansion in straining flows ($\lambda > 0$ for $\Phi\ll -\omega^4$), and the asymptotic variation of expansion or contraction rate at high amplitude oscillations ($\lambda \propto |\Phi|^{1/4}$ for $|\Phi| \gg 1$). 
These two models also fail to predict particle crossovers that occur in strong-straining flows where $\imag\lambda \ne 0$.

\section{Three-dimensional isotropic turbulence: extension to multimodal excitation} \label{HIT}
In a physically realistic turbulent flow, excitation is not at a single frequency but involves a continuous range of frequencies.
To analyze a multimodal excitation, all modes that appear in Eq.~\eqref{lambda} must be retained. 
With the transformation introduced in Eq.~\eqref{gam_def},  Eq.~\eqref{lambda} can be expressed as
\begin{equation}
\frac{n}{4}\left(\gamma^2 - 1\right) - \iinf \frac{ \tilde \rho^{\rm Q}(\omega; {\rm St})}{\gamma^2 + \omega^2} \dd \omega=0,
\label{gam_gen}
\end{equation}
which provides a generic relationship for $\gamma^2$ in the form of an eigenvalue problem.
An iterative approach can be adopted for computing $\gamma^2$ from this equation, as the evaluation of the integral requires prior knowledge of $\gamma^2$.

Under certain conditions, it is possible to find an explicit closed-form relationship for $\gamma$ based on Eq.~\eqref{gam_gen}.
One such scenario is when the explicit form of $\tilde \rho^{\rm Q}(\omega; {\rm St})$ is known, as was the case in Section~\ref{unimodal}.
If the analytical form of $\tilde \rho^{\rm Q}(\omega)$ were to be available and $\tilde \rho^{\rm Q} \to 0$ as $|\omega| \to \infty$, Cauchy's integral formula could be employed to express
\begin{equation}
\iinf \frac{ \tilde \rho^{\rm Q}(\omega; {\rm St})}{\gamma^2 + \omega^2} \dd \omega = \frac{\pi}{\gamma} \tilde \rho^{\rm Q}(\hat i \gamma; {\rm St}).
\label{cauchy}
\end{equation}
Depending on the form of $\tilde \rho^{\rm Q}(\hat i \gamma; {\rm St})$, an explicit relationship for $\gamma$ can be obtained from this expression.

Another possible scenario that may arise is a design problem formulated as finding a specific $\tilde \rho^{\rm Q}(\omega; {\rm St})$ when a desirable $\gamma({\rm St})$ is given.
Designing a hydrodynamic particle separator by enhancing the clustering of a particular class of particles in a polydisperse distribution is an instance of such a scenario.
The present formulation can be instrumental in solving this inverse problem by expressing Eq.~\eqref{gam_gen} as a Fredholm integral equation of the first kind with a kernel function $(\gamma^2+\omega^2)^{-1}$ \cite{hildebrand2012methods}. 

\subsection{$\tilde \rho^{\rm Q}$ in a turbulent flow}
In a nutshell, $\tilde \rho^{\rm Q}$ determines at what frequencies different gradients in the fluid velocity field oscillate as seen by the particle.
Analytical exploration of $\tilde \rho^{\rm Q}$ is a non-trivial task for an arbitrary flow.  
In general, fluid flows are too complex to be solved analytically and expressed in a closed-form solution. 
This statement is particularly true about turbulence that is typically studied experimentally or numerically when a detailed solution is needed. 
What makes the analytical calculation of $\tilde \rho^{\rm Q}$ even more challenging in such flows is its dependence on the particle trajectory. 
Besides having an explicit relationship for the velocity gradient at all points in space and time, one needs to know how particles preferentially sample the flow to compute $\tilde \rho^{\rm Q}$. 
Despite all these complexities, analytical modeling of $\tilde \rho^{\rm Q}$ in an approximate form could be a feasible task to be accomplished by future investigations.
An essential component in such an investigation would be a model for the velocity gradients along particle trajectories. 
Such an effort has been undertaken in the context of large-eddy simulations~\cite{johnson2018predicting} for tracers and needs to be extended to the inertial particles.
In this study, we do not attempt to model $\tilde \rho^{\rm Q}$, but rather compute it directly from a numerical simulation.
This way, the accuracy of the present analysis can be evaluated independently of the accuracy of the model used for $\tilde \rho^{\rm Q}$.

Although the results presented in this section are based on the accurate quantification of $\tilde \rho^{\rm Q}$, in Section~\ref{1D_SR} we will show that such an accurate knowledge is not necessary for a qualitative prediction of $\mathcal C$.
Even with a minimal knowledge of the flow, i.e., the fact that particles preferentially sample the straining region of the flow or $\langle Q_\eta \rangle < 0$, the present analysis can predict the existance of a dip in $\langle \mathcal C_\eta \rangle$ at $\rm {St}\approx 1$ and $\langle \mathcal C_\eta \rangle \to 0$ as $\rm {St} \to 0$ or $\infty$.
This lack of sensitivity to the form of $\tilde \rho^{\rm Q}$ shows that the clustering phenomenon is rather universal, reaffirming our intuition that the clustering ought to occur primarily at $\rm {St} \approx 1$ regardless of the details of the flow field. 

To generate the background flow, we perform direct numerical simulation of a triply periodic homogeneous isotropic turbulence using an in-house solver with a specialized linear solver \cite{esmaily2018scalable}. 
A second-order spatial discretization on a $256^3$ grid and 4$^{\rm th}$ order Runge-Kutta time integration scheme are employed. 
Stationary turbulence is maintained by adding a forcing term to the momentum equation that is proportional to the velocity \cite{rosales2005linear}.
The forcing term is dynamically computed at each time step to prevent fluctuation of $\tau_\eta$ and thus St \cite{esmaily2016CPT,bassenne2016constant}. 
The maximum deviation of $\tau_\eta$ from the target Kolmogorov time scale is 0.3\%. 
The Reynolds number based on the Taylor micro-scale is Re$_\lambda=100$.
Special care has been taken in interpolating quantities at the location of particles from the Eulerian grid.
In particular, the interpolation scheme is designed to correctly translate the incompressibility condition to the Lagrangian velocity gradient tensor. 
Additionally, the Lagrangian gradients are kept $C^0$ continuous by interpolating from a pre-constructed continuous Eulerian field \cite{esmaily2016CPT}.

\begin{figure}
\begin{center}
\includegraphics[width=1.0\textwidth]{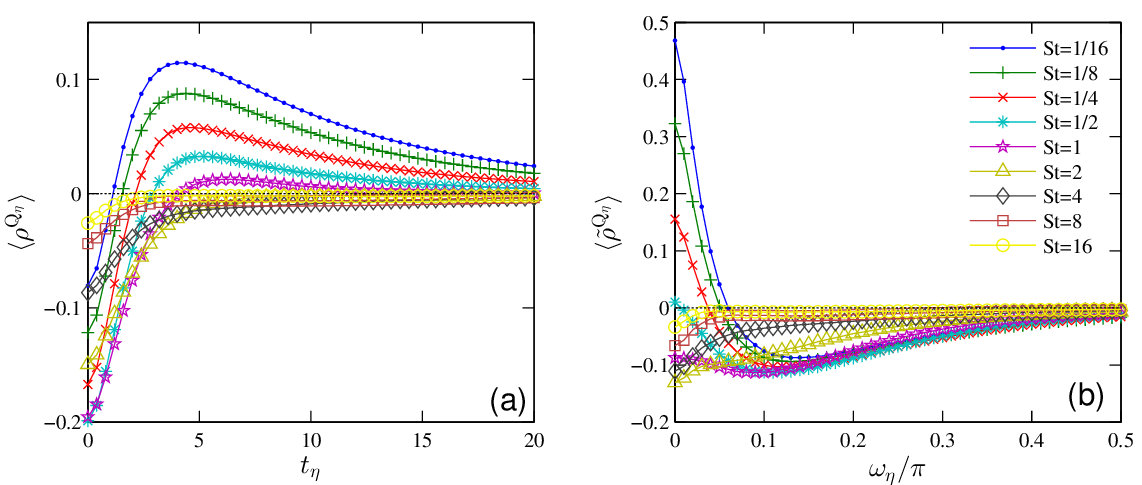}
\caption{The ensemble average of $\rho^{\rm Q_\eta}$ (a) and its Fourier transformation (b) -- defined in Eq. \eqref{Q_def} -- computed along the particle trajectories with different Stokes numbers in an isotropic turbulent flow at ${\rm Re}_\lambda = 100$ (colors online).}
\label{fig:rQt3}
\end{center}
\end{figure}

The particles trajectory is computed using Eq.~\eqref{stokes}.
A dilute mixture is considered so that the effect of particles on the fluid can be neglected, thereby eliminating potential errors that can be introduced by two-way coupling forces \cite{horwitz2016accurate, ireland_desjardins_2016, esmaily2018correction}.
113 classes of particles are considered in total with ${\rm St} = 2^{p/8}$, $p\in\{-32,\ldots,80\}$.
At each Stokes number, approximately $10^4$ randomly seeded particles were simulated for several large eddy turnover time to allow development of clusters.
Starting with this time-evolved distribution, we record the velocity gradient tensor at the position of each particle for 700$\tau_{\eta}$ with $0.1\tau_\eta$ intervals.
Based on $\nabla \bl u_\eta(t)$, $\tilde \rho^{\rm Q_\eta}$ is computed using Eq.~\eqref{Q_def} at each St.
The number of particles and the integration period are verified to be sufficient for achieving statistical convergence.
The results of these calculations at few Stokes numbers are shown in Figure~\ref{fig:rQt3}.

In a turbulent flow, in contrast to the previous case study, $\nabla \bl u_\eta$ has full rank, and $\|\bl S_\eta\|$ and $\|\bl \Omega_\eta\|$ are nonzero simultaneously.
These parameters, computed along the trajectory of particles, exhibit complex behaviors due to their dependence on St, caused by the preferential sampling of the flow field by the particles.
These dependencies are briefly mentioned here for the sake of completeness and discussed in detail in \cite{esmaily2016CPT}.
In the homogeneous turbulence under consideration, $\rho^{\rm S_\eta}(t;{\rm St})$ and $\rho^{\rm \Omega_\eta}(t;{\rm St})$ (the norm of the autocorrelation function of the strain- and rotation-rate tensors, respective) are both strictly positive. 
$\langle \rho^{\rm S_\eta} \rangle$ is relatively independent of St and exponentially decays with time, which is analogous to what can be observed with particles in a random straining flow.
$\langle \rho^{\rm \Omega_\eta} \rangle$, on the other hand, varies significantly versus St.
As a result, $\langle \rho^{\rm Q_\eta}\rangle = \langle \rho^{\rm \Omega_\eta}\rangle - \langle\rho^{\rm S_\eta}\rangle$ strongly depends on St.

For ${\rm St} \lesssim 1$, $\langle \rho^{\rm Q_\eta} \rangle$ undergoes an increasing-decreasing trend in time.
$\langle \rho^{\rm Q_\eta} \rangle$ being negative at $t_\eta \ll 1$ and ${\rm St}\ll 1$ is due to the smaller value of $\langle \rho^{\rm \Omega_\eta} \rangle$.
Hence, particles with small St tend to centrifuge out of rotational regions with the short time constant and follow slow vortical features since $\langle \tilde \rho^{\rm Q_\eta}\rangle > 0$ at $\omega_\eta \ll 1$ and ${\rm St} < 1$ (Figure \ref{fig:rQt3}).

For ${\rm St} \gg 1$, particles are not responsive to the velocity fluctuations and follow a trajectory that is uncorrelated with the flow.
As a result, particles distribute uniformly in space and the Lagrangian and Eulerian statistics become almost identical.
Additionally, one can show that the Eulerian strain-rate and rotation-rate autocorrelation functions are equal in a periodic domain.
Therefore, $\langle \rho^{\rm S_\eta} \rangle$ and $\langle \rho^{\rm \Omega_\eta} \rangle$ converge to the same value, leading to $\langle \rho^{\rm Q_\eta}\rangle \to 0$ and $\langle \tilde \rho^{\rm Q_\eta}\rangle \to 0$ as ${\rm St} \to \infty$. 

In Section \ref{numerical}, we demonstrated that the particle clouds only contract in a straining regime as a rotating regime only leads to cloud expansion. 
Thus, in turbulence, where regions of higher rotation-rate and strain-rate coexist in space, particles tend to accumulate in regions of higher strain-rate. 
The preferential concentration of particles in the straining regions, which occurs at all St, is supported by the dominance of $\langle \rho^{\rm S_\eta}\rangle (t=0)$ over $\langle \rho^{\rm \Omega_\eta} \rangle (t=0)$.
This dominance is most noticeable at ${\rm St} \approx 1$, where $\langle \rho^{\rm Q_\eta}\rangle(0)$ has an extremum.
At large and small Stokes numbers, on the other hand, $\langle \rho^{\rm Q_\eta}\rangle(0)$ asymptotes to zero as $\langle\rho^{\rm S_\eta}\rangle(t=0)$ and $\langle \rho^{\rm \Omega_\eta}\rangle(t=0)$ reach a balance. 

\subsection{Clustering statistics} \label{3d-cluster}
\begin{figure}
\begin{center}
\includegraphics[width=1.0\textwidth]{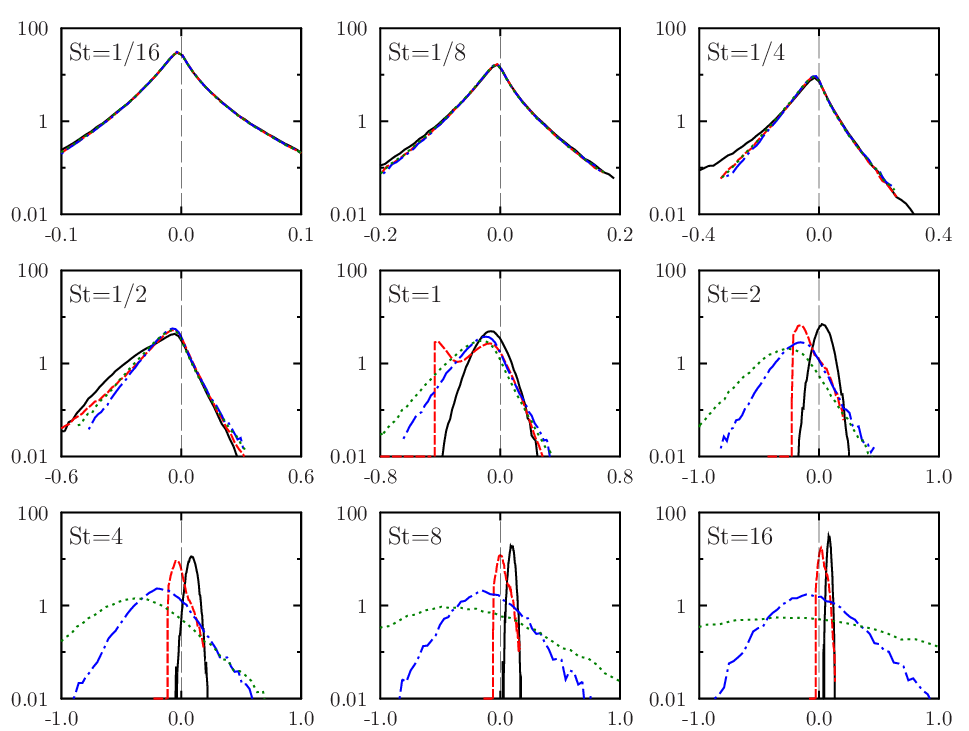}
\caption{The PDF of finite-time contraction-rate $\mathcal C^t_\eta$ at different Stokes numbers obtained based on RM (dotted green), SL (dashed-dotted blue), the present analysis (dashed red), and the reference numerical results (solid black) from the three-dimensional isotropic turbulence at ${\rm Re}_\lambda=100$.
While all models provide good approximation at low St, only the present analysis agrees with the reference results at high St (colors online).}
\label{fig:C3_pdf}
\end{center}
\end{figure}

From $\tilde \rho^{\rm Q_\eta}$, $\mathcal C^t_\eta$ is computed for RM, SL, and the present analysis.
The method described in Section \ref{numerical} is employed to compute the reference quantities. 
To improve accuracy, computations are performed with 12 time steps between consecutive records of $\nabla \bl u_\eta$, i.e. $\Delta t=\tau_\eta/120$. 
To prevent ill-conditioning of $\bl A$ in Eq.~\eqref{A_trans}, the time integration period is limited to 12.5$\tau$ when $\tau < 56$, whereas the entire interval of 700$\tau_\eta$ is considered for $\tau \ge 56$. 
This division to sub-intervals is uniformly applied to the computation of the input to the models and also the calculation of the reference results to ensure a one-to-one comparison between the two.
To construct the PDF of $\mathcal C^t_\eta$, each sub-interval associated with each $10^4$ particles is treated as an independent ensemble (Figure \ref{fig:C3_pdf}).
The overall trend is similar to what was observed in Section~\ref{comp} for the one-dimensional case with the present analysis being the closest to the reference followed by SL and RM.
The only exception is ${\rm St}=1$ where the PDF from the present analysis is skewed and shows a second unphysical peak at $\mathcal C^t_\eta \approx -0.5$.

\begin{figure}
\begin{center}
\includegraphics[width=1.0\textwidth]{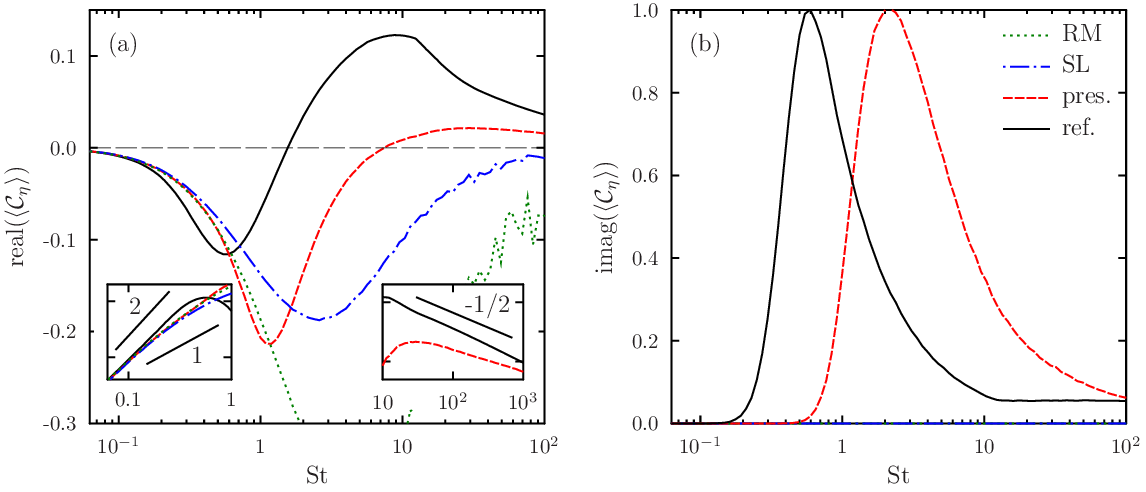}
\caption{The rate of expansion or contraction (a) and crossover (b) of particle clouds as a function of Stokes number based on RM (dotted green), SL (dashed-dotted blue), the present analysis (dashed red), and the reference numerical simulations (solid black).
The underlying flow is a three-dimensional stationary isotropic turbulence at ${\rm Re}_\lambda=100$.
The crossover rates are normalized by their maximum value. 
The left and right insets are the same plot for ${\rm St}\le1$ and  ${\rm St}\ge10$ in a logarithmic scale, respectively.
Lines with a slope of 1, 2, and $-1/2$ are shown for reference (colors online).}
\label{fig:C3}
\end{center}
\end{figure}

The ensemble-averaged of $\mathcal C^t_\eta$ is computed from the PDFs (Figure \ref{fig:C3}).
Since RM and SL are a linear function of $\langle \tilde \rho^{\rm Q_\eta}\rangle$, these ensemble-averaged quantities can be computed directly from the results shown in Figure \ref{fig:rQt3}.
For the present analysis, however, $\langle \mathcal C_\eta(\tilde \rho^{\rm Q_\eta}) \rangle$ is slightly different from $\mathcal C_\eta(\langle \tilde \rho^{\rm Q_\eta}\rangle)$. 
Their difference depends on the integration period $t_\eta$ and asymptotes to zero as $t_\eta \to \infty$.

The accuracy of all three models in predicting the reference results is similar to what was observed for the one-dimensional cases. 
All models collapse with the reference for ${\rm St} \ll 1$, whereas for ${\rm St} \gg 1$, their prediction widely varies. 
Among the three models, only the present analysis captures the expansion of clouds at ${\rm St} \gg 1$. 
$\real{\langle \mathcal C_\eta \rangle} > 0$ is predicted at ${\rm St} \gg 1$ despite the fact that $\langle \tilde \rho^{\rm Q_\eta}\rangle < 0$ at all frequencies (Figure \ref{fig:rQt3}). 
Prediction of expansion in a straining regime stems from the nonlinear behavior of $\lambda(\Phi)$ that appeared as $\real\lambda > 0$ for $\Phi < -(\omega^4 + 10\omega^2 + 0)/16$ in Figure \ref{fig:C1a}. 

Despite capturing the overall trend, the present analysis is not in full quantitative agreement with the reference at high St. 
The disagreement can be attributed to the assumption of single $\lambda$ in Eq.~\eqref{F_guess}, where a full rank matrix was replaced with a diagonal matrix\footnote{Our earlier argument that all $\lambda_i$ are the same in an isotropic flow relies on a sufficiently long sampling period to diminish statistical differences between $\lambda_i$. Here, however, the sampling period is limited to 12.5$\tau$ to prevent ill-conditioning of $\bl A$, resulting in 3 distinct $\lambda_i$ for each ensemble, which often are significantly different.}, and excluding sub-harmonics from our analysis when deriving Eq.~\eqref{F_2_i}.

There are two predictions that are unique to the present analysis and in agreement with the reference results.
The first is the asymptotic behavior of $\langle \mathcal C_\eta \rangle$ as ${\rm St} \to \infty$ and the second is the sequence of the onset of crossover and dispersion. 
At very large Stokes number, $\langle \mathcal C_\eta\rangle \propto \rm{St}^{-1/2}$ (the right inset in Figure~\ref{fig:C3}), which is in agreement with the prediction of the present analysis (Figure~\ref{fig:p-space}). 
Additionally, the present model predicted crossovers to occur at a Stokes number that precedes the Stokes number at which clustering is transitioned to dispersion regime. 
Based on Figure~\ref{fig:C3}, the onset of crossover and dispersion occur at $\rm St\approx 0.18$ and $\rm St\approx 1.6$, respectively, following the predicted sequence.
The ratio of two Stokes number, however, does not quantitatively agree with our model ($\approx9$ versus 3). 

\begin{figure}
\begin{center}
\includegraphics[width=0.5\textwidth]{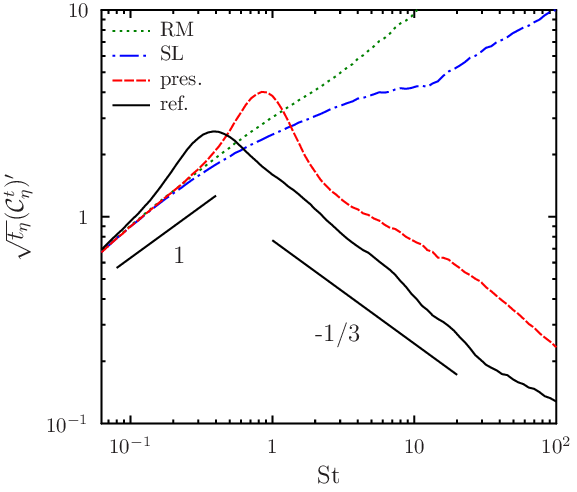}
\caption{The normalized standard deviation of the rate of contraction as a function of Stokes number, obtained from RM (dotted green), SL (dashed-dotted blue), the present analysis (dashed red), and the reference numerical simulations (solid black).
Results corresponds to the three-dimensional isotropic turbulence at ${\rm Re}_\lambda=100$ (colors online).}
\label{fig:C3rms}
\end{center}
\end{figure}

To show the effect of turbulence intermittency on the particle clustering, we computed the second moment of $\mathcal C^t_\eta$ (Figure \eqref{fig:C3rms}).
The plotted moments are normalized by $\sqrt{t_\eta}$ to ensure their independence from sampling period $t$.
Among the available models, the present analysis provides the best estimation for $(\mathcal C_\eta^t)^{\prime}$.
All models collapse with the direct computations at the limit of small St and deviate from it as St increases.
For ${\rm St} > 1$, RM and SL predictions linearly increase with St while the present analysis remains bounded.
A slope of $-1/3$ is observed in this regime of Stokes number, indicating a reduction in sensitivity of particle to turbulence fluctuations, which can be explained by their higher inertia and smoother trajectories~\cite{kadoch2016multiscale}.

\subsection{Relevance of the one-dimensional model problem} \label{1D_SR}
In section \ref{unimodal}, we considered a very simple flow that was one-dimensional and oscillated at a single frequency. 
The results of that analysis were primarily presented as a function $\Phi$ and $\omega$ that characterize the amplitude and frequency of oscillations, respectively, when normalized based on the particle relation time $\tau$. 
In Section~\ref{St-renorm}, we showed that those results become $\rm St$-dependent if they are re-normalized based on the flow time-scale $\tau_\eta$. 
Later in Section~\ref{3d-cluster}, we applied a more general version of the same analysis to a three-dimensional isotropic turbulent flow and presented the results that were also normalized based on the flow time-scale $\tau_\eta$.
Our goal in this section is to find how the result of these two cases compare in terms of St and whether the simple one-dimensional flow can adequately lead to an understanding of the clustering phenomenon in the three-dimensional turbulent flow. 

\begin{figure}
\begin{center}
\includegraphics[width=1.0\textwidth]{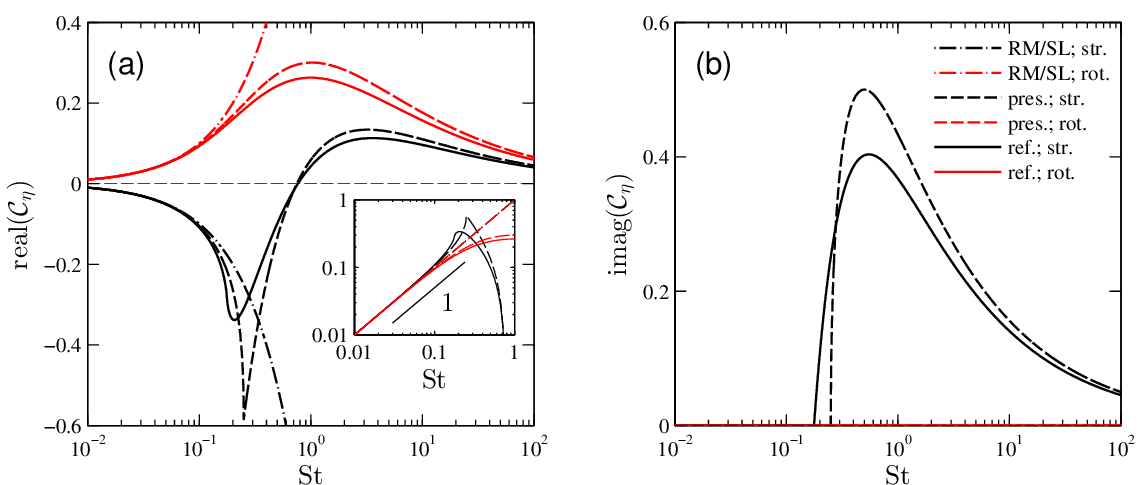}
\caption{The rate of contraction or expansion (a) and crossover (b) of particle clouds as a function of St in a low frequency oscillatory one-dimensional flow.
RM and SL (dash-dotted) and the present analysis (dashed) are compared against the reference numerical computation (solid) for a straining (black) and rotating (red) flow.
Inset: The same plot in the logarithmic scale for ${\rm St} \le 1$ with the curves associated with the straining flow inverted (colors online).}
\label{fig:C1}
\end{center}
\end{figure}

Let us revisit the one-dimensional flow of Section~\ref{unimodal} in the regime of $\omega \ll 1$.
The condition of $\omega\ll 1$ corresponds to a flow oscillating at a frequency much lower than the inverse of particle relaxation time $\tau$.
Based on the discussion of Section~\ref{St-renorm}, $\lambda_\eta$ for this flow can be computed from Eq.~\eqref{lambda_1D_St} with $\omega_\eta \approx 0$.
Since $\lambda_\eta$ can be replaced by $\mathcal C_\eta$ for a one-dimensional flow, Eq.~\eqref{lambda_1D_St} can be written as
\begin{equation}
\mathcal C_\eta = -\frac{1}{{2\rm St}} + \frac{1}{4{\rm St}}\sqrt{2+2\sqrt{1 + 16{\rm St}^2\Phi_\eta}}, \;\;\; \Phi_\eta=\pm 1,
\label{C1_eta}
\end{equation}
where $\Phi_\eta = +1$ and $-1$ represent a rotating and straining flow, respectively\footnote{Note that $\Phi_\eta = \overline{\|\bl \Omega_\eta\|^2 - \|\bl S_\eta\|^2}$ (see Eq.~\eqref{1Du_def}). Thus, $\Phi_\eta=\pm 1$ since $\tau_\eta$ is taken as $\|\bl S_{\rm d}\|^{-1}$ or $\|\bl \Omega_{\rm d}\|^{-1}$, where $\bl S_{\rm d}$ and $\bl \Omega_{\rm d}$ are the dimensional strain- and rotation-rate tensors, respectively.}.
For this one-dimensional flow in which $\omega_\eta \approx 0$, the prediction of SL collapses with that of RM to $\mathcal C_\eta = \pm {\rm St}$. 
The reference results can also be obtained by following the procedure of Section~\ref{numerical} using $\omega_\eta = 10^{-4}$. 
The result of all these calculation is plotted in Figure \ref{fig:C1}, where $\imag{\mathcal C_\eta}$ is also shown as the measure of particle crossover frequency. 

As mentioned earlier, the accumulation of particles in the straining region of a turbulent flow leads to $\langle \|\bl S_\eta \| \rangle > \langle \|\bl \Omega_\eta \| \rangle$.
As a result, the behavior of $\mathcal C_\eta({\rm St})$ in the isotropic turbulence (Figure \ref{fig:C3}) should be compared against the one-dimensional straining flow (black curves in Figure \ref{fig:C1}) rather than the rotating flow. 
One can verify that there is a one-to-one correspondence between the two cases. 
Increasing St from zero, $\real{\mathcal C_\eta}$ decreases till the onset of crossover. 
The trend is reversed once ${\rm imag}(\mathcal C_\eta) \ne 0$ up to a Stokes number at which $\mathcal C_\eta > 0$.
For larger St, $\real{\mathcal C_\eta}$ changes non-monotonically with $\mathcal C_\eta \propto {\rm St}^{-1/2}$ as ${\rm St} \to \infty$ (not shown in Figure~\ref{fig:C1} explicitly).
Such a close similarity between one- and three-dimensional cases provides a window from which one can view how clustering phenomenon unfolds as the Stokes number changes in a flow.
We provide this description for the simpler one-dimensional case that is easier to understand, however, one can extend the following arguments to the turbulent flow that in essence is three oscillatory one-dimensional straining or rotating flows acting on a particle cloud (along the principal directions of the velocity gradient tensor) as it traverses the flow.

To better understand the behavior of particles in the straining flow (black curves in Figure \ref{fig:C1}), it is necessary to distinguish between two regimes in which $\real{\mathcal C_\eta}$ decreases and increases with St. 
In the first regime occurring at ${\rm St} < \sqrt{1/32}$\footnote{This transition is predicted to occur at ${\rm St} \approx \sqrt{1/16}$ rather than $\sqrt{1/32}$ by the present analysis despite the fact that the corresponding extremum of $\real{\lambda}$ was correctly predicted in Section~\ref{HIT}. A closer examination shows that the extremum of $\real{\mathcal C_\eta}$ is shifted to higher St due to the re-normalization of Eq.~\eqref{C1_eta} and under-prediction of $\real\lambda$ by our analysis (note the extremum of $\real{\mathcal C_\eta({\rm St})}$ occurs at ${\rm St} = \sqrt{-\Phi}$ with $\Phi$ satisfying $\real{\lambda(\Phi)} = 2\Phi [{\rm d} (\real\lambda)/{\rm d} \Phi]$).}, $\imag{\mathcal C_\eta} = 0$ and $\real{\mathcal C_\eta}$ decrease almost linearly versus St.
No particle crossover occurs in this regime and the increase in the particle inertia is met with proportionally stronger slippage, leading to a faster rate at which particles get close to each other ($\{\real \lambda < 0 , \; \imag \lambda = 0\}$ in Figure \ref{fig:two_prt}). 
Further increase in St leads to the second regime, in which particles have enough inertia to cross over each other ($\imag \lambda \ne 0$ in Figure \ref{fig:two_prt}).
The maximum rate of convergence is obtained at the onset of crossovers before particles begin to overshoot each other.
For ${\rm St}> \sqrt{1/32}$, the relative velocity of particles at the moment of crossover increases with St, such that at ${\rm St} \ge 3/4$ their mean distance, rather than decreasing, begins to increase over time. 
As St is increased beyond ${\rm St} \approx 3.6$, the rate at which particles diverge decreases, i.e., ${\rm d}(\real{\mathcal C_\eta})/{\rm d St} < 0$ for ${\rm St} > 3.6$.
Due to the high inertia of particles at this limit, particles hardly respond to the oscillations of the underlying flow and as a result tend to maintain their initial position, lowering their divergence rate.
Another consequence of having very inertial particles is fewer incidents of crossovers that leads to a decrease in $\imag{\mathcal C_\eta}$.

In contrast to the one-dimensional straining flow, which produced a clustering phenomenon analogous to the isotropic turbulent flow, it is hard to find a real-world example that corresponds to the one-dimensional rotating flow (red curves in Figure~\ref{fig:C1}).
The reason is that the most commonly studied spatially and temporally oscillating particle-laden flows are turbulent.
These flows contain regions in space where $\|\bl S\| < \|\bl \Omega\|$ and $\|\bl S\| > \|\bl \Omega\|$. 
Inertial particles in these flow tend to accumulate in regions where $\|\bl S\| > \|\bl \Omega\|$, thus leading to a behavior that is similar to that of the straining flow. 
If one were to identify a spatially and temporally oscillating flow that on average exposes particles to a higher rotation- than strain-rate (i.e., $Q>0$), then we may observe a behavior similar to that of the one-dimensional rotating flow in Figure~\ref{fig:C1}.
The behavior of particles in such rotating flow will be less complicated as there is no particle crossover. 
The distance between particles always increases over time at $\mathcal O({\rm St})$ rate at ${\rm St} \ll 1$ and $\mathcal O({\rm St}^{-1/2})$ at ${\rm St} \gg 1$ (if dependence of $Q_\eta$ on St is neglected). 
The asymptotic behavior of particles at these two limits will remain similar to the straining flow. 
At small St, particles follow fluid tracers and their distance barely changes over time. 
At high St, particles barely respond to the underlying flow oscillations and maintain their initial position\footnote{Since $\real{\mathcal C_\eta({\rm St})}$ is linear at low St, the magnitude of $\real{\mathcal C_\eta}$ for straining and rotating flows is equal up to the leading order term. 
At large St, however, the leading order terms have a similar exponent, i.e. $\real{\mathcal C_\eta({\rm St})} \propto {\rm St^{-1/2}}$, but different magnitude with the rotating flow leading to a larger magnitude (Figure~\ref{fig:C1}).}.

There are also some differences between the results of the one-dimensional straining flow and the three-dimensional turbulence.
In the turbulent flow, the maximum contraction rate occurs at a higher St and is less significant (Figures \ref{fig:C3} and \ref{fig:C1}).
This weaker clustering can be attributed to the presence of rotating regions that are absent in the one-dimensional case.
Additionally, the asymptotic behavior of $\real{\mathcal C_\eta}$ at ${\rm St} \ll 1$ is linear for the one-dimensional case, whereas it is superlinear for the turbulent flow.
Since $\real{\langle \mathcal C_\eta \rangle} \approx {\rm St} \langle Q_\eta \rangle$ at small St, this difference is a result of the behavior of $\langle Q_\eta \rangle$ versus St.
$\langle Q_\eta \rangle$ was independent of St in the one-dimensional case, hence the linear rate, whereas it is proportional to St in the turbulent case\footnote{The rate at which $\langle Q_\eta \rangle$ grows versus St in an isotropic turbulent flow can be an artifact of the periodic boundary condition imposed for computational consideration. This potential artifact caused by the spatial confinement must be removed in the future for a more realistic assessment of the asymptotic behavior of $\real{\langle \mathcal C_\eta \rangle}$ at small St in a turbulent flow.}, hence the superlinear rate.
Note the differences in the magnitude of $\imag{\mathcal C_\eta}$ is immaterial since the three-dimensional results are normalized by their maxima. 
Finally, the Stokes at which clustering is maximized coincides with the onset of crossovers for the one-dimensional case, whereas it occurs after the onset of crossovers for the turbulence case.
This difference could be a result of the turbulence intermittency as $\imag{\mathcal C_\eta}\ne 0$ for a fraction of ensembles leads to $\imag{\langle \mathcal C_\eta\rangle}\ne 0$, thus shifting the onset to smaller St where only the trajectories of a few particles cross. 

In overall, the one-to-one correspondence between the results of the one- and three-dimensional cases is remarkable (Figure~\ref{fig:C3} versus black curves in Figure~\ref{fig:C1}). 
The former is a simple flow expressed by Eq.~\eqref{1Du_def} oscillating at a single frequency, while the latter involves a multiscale three-dimensional chaotic flow with gradients oscillating at a continuous spectrum of frequencies. 
Such a close comparison between the two is a testimony to the fundamental significance and relevance of the one-dimensional model problem.
It confirms our earlier hypothesis that the clustering phenomenon and how it varies versus St is primarily determined based on the governing equations of the motion of particles rather than the detailed structure of the underlying flow.

\section{Conclusions}
We derived a solution (Eq.~\eqref{lambda}) for the Lyapunov exponents of inertial particles subjected to oscillatory fluid motion.
Our analysis is aimed at predicting the rate of expansion or contraction of clouds of inertial particles, and also their crossovers frequency. 
We employed the sum of the Lyapunov exponents, i.e., the rate of change of volume of a cloud of particles in three dimensions, to characterize regimes of preferential concentration. 
We showed that our solution is more general and reproduces the pre-established models in the literature (\cite{maxey1987gravitational} and \cite{esmaily2016CPT}).
Consistent with the previous models, the only flow-related parameter that appears in our model is the difference between the spectrum of rotation and strain rate tensors norm, viz. a closely related parameter to the Q-criterion, underscoring its fundamental role in clustering of inertial particles. 
We employed a canonical setup with unimodal excitation to investigate the behavior of the Lyapunov exponent under a wide range of flow conditions. 
Only the expansion with no crossovers was observed in a rotating regime, whereas both the contraction and expansion with the possibility of crossovers was observed in a straining regime. 
In a straining regime, the expansion and crossover occur for a sufficiently large oscillation amplitude. 
Additionally, a $-1/2$ bound on the rate of contraction (normalized by the particle relaxation time) was found. 
Our analysis also showed the Lyapunov exponent is linearly proportional to the Q-criterion at low oscillation amplitude and its power of $1/4$ at high oscillation amplitude.
These observations, which are confirmed by the numerical simulations, are not captured by the other models.
Other available models capture only the linear regime, where the oscillation amplitude is small. 
Discrepancies were also observed between our analysis and the reference results. 
Neglecting the contribution of the sub- and super-harmonics in our formulation are deemed to be the primary sources of discrepancies. 
Accounting for higher-order terms and extending the present analysis to statistically anisotropic flows remains as topics for future studies. 

Following this canonical setting, we extended our analysis to complex multi-dimensional flows, in which a continuous range of frequencies is present. 
We considered a three-dimensional isotropic forced turbulence for validation of our analysis. 
Despite the added complexity, this case produced results analogous to that of the one-dimensional straining regime with unimodal excitation. 
In all cases, the contraction rate was proportional to St at small St with an extremum around the onset of crossovers. 
For larger St, the rate of contraction decreases till net expansion is observed at ${\rm St} \approx 1$.
Beyond that St, the rate of expansion reaches a maximum and then asymptotes to zero proportional to $\rm St^{-1/2}$ as ${\rm St}\to \infty$. 
While all models correctly capture the linear trend at low St, only the present analysis provides a good prediction of the subsequent nonlinear behaviors at higher Stokes numbers.
Additionally, only our analysis captures the occurrence of particle crossovers at high St and also the non-monotonic variation of the standard deviation of the rate of expansion or contraction versus St. 

\section*{acknowledgments}
We gratefully thank Prof. Andreas Acrivos for his extensive effort in reading this manuscript and providing detailed comments, which have been quite instrumental in clarification of the introduced concepts.
This work was supported by the United States Department of Energy under the Predictive Science Academic Alliance Program 2 (PSAAP2) at Stanford University.

\appendix
\section{Derivation of Eq.~\eqref{C_def}} \label{der_C_Ct}
To relate $\mathcal C$ to $\mathcal C^t$, the Eulerian form of Eq.~\eqref{RM} must be expressed in terms of Lagrangian quantities. 
Since $\nabla$ operator can be expressed as $\partial /\partial x_i = ( \partial /\partial X_j )(\partial X_j/\partial x_i)$, we have
\begin{equation}
\nabla \cdot \dot{\bl x} = \frac{\partial \dot x_i}{\partial X_j} \frac{\partial X_j}{\partial x_i}  =  \dot J_{ij} J^{-1}_{ji}.
\label{a_1}
\end{equation}
From Jacobi's formula
\begin{equation}
\dot J_{ij} J^{-1}_{ji} = \det{\bl J } ^{-1} \ddt {[\det{\bl J}]},
\label{Jacobi}
\end{equation}
and as a result
\begin{equation}
\nabla \cdot \dot{\bl x} = \frac{\dd \left(\ln[ \det{\bl J} ]\right)}{\dd t}.
\label{a_2}
\end{equation}
From Eqs.~\eqref{a_2} and \eqref{Ct_def}
\begin{equation}
\frac{\dd \left(t \mathcal C^t \right)}{\dd t} = \nabla \cdot \dot{\bl x},
\label{a_3}
\end{equation}
which in combination with Eq.~\eqref{RM} gives
\begin{equation}
   \mathcal C = \overline{\frac{\dd \left(t \mathcal C^t \right)}{\dd t}} = \lim_{t\to \infty} \frac{1}{t} \int_0^t \frac{\dd \left(t^\prime \mathcal C^{t^\prime} \right)}{\dd t^\prime} \dd t^\prime = \lim_{t\to \infty} \mathcal C^t,
\label{a_4}
\end{equation}
completing the derivation.  

\section{Derivation of Eq.~\eqref{C_cal}} \label{app:eqC}
To relate $\mathcal C$ to $\bl F$, we start from the Jacobi formula in Eq.~\eqref{Jacobi_F} that can expressed as
\begin{equation}
\tr(\bl F) =  \ddt {\left( \ln[ \det{\bl J(t)} ]\right)}.
\label{app2_1}
\end{equation}
Averaging Eq.~\eqref{app2_1} over time produces
\begin{equation}
\overline{\tr(\bl F)} =  \overline{\ddt {\left( \ln [\det{\bl J(t)}] \right) }} = \lim_{t\to \infty} \frac{1}{t} \int_0^{t} {\frac{\dd \left( \ln [\det{\bl J(t^\prime)}] \right)}{\dd t^\prime}} \dd t^\prime.
\label{app2_2}
\end{equation}
Thus
\begin{equation}
\overline{\tr(\bl F)} = \lim_{t\to \infty} \frac{\ln [\det{\bl J(t)}] - \ln[ \det{\bl J(0)}]}{t}.
\label{app2_3}
\end{equation}
From Eq.~\eqref{Ct_def} and $\det{\bl J(0)} = 0$, the above equation can be simplified to
\begin{equation}
\overline{\tr(\bl F)} = \lim_{t\to \infty} \mathcal C^t.
\label{app2_4}
\end{equation}
Based on Eq.~\eqref{C_def}, the term on the right-hand side of Eq.~\eqref{app2_4} is $\mathcal C$, thus completing the proof. 

\section{Higher order expansions} \label{app:HOE}
To explain the difference between the analytical prediction and the numerical result observed in Section \ref{numerical}, we need to revisit the underlying assumptions of Eq.~\eqref{lambda_1D}.
Equation \eqref{lambda_1D} is an exact solution of Eq.~\eqref{lambda} that was obtained from Eq.~\eqref{F_eq} when the higher order terms in Eq.~\eqref{F_1_i} are neglected.
To account for those higher order effects, we consider
\begin{equation}
\bl F = \lambda \bl I + \sum_\omega \bl \epsilon_1(\omega) e^{\hat i \omega t} + \sum_\omega \bl \epsilon_2(\omega) e^{2 \hat i \omega t} + \cdots.
\label{F_guess_app}
\end{equation}
as an asymptotic form of $\bl F$, where $\bl \epsilon_1$ corresponds to $\bl \Psi$, which is the first order solution obtained in Section \ref{ana_der}.
In this expression, the higher order terms, i.e. $\bl \epsilon_i$ for $i>1$, can be computed such that the second summation in Eq.~\eqref{F_1_i} is represented more accurately. 
The addition of higher order terms modifies $\lambda$ in two ways. 
The first is to interact with the lower order terms and alter their amplitude.
The second is to directly contribute to $\lambda$ through contraction of $\bl \epsilon_i(\omega)$ and $\bl \epsilon_i(-\omega)$.
In either case, $\| \bl \epsilon_i \|$ provides a measure of the significance of those higher order terms. 
Therefore, we evaluate the importance of these higher order terms by investigating the decay rate of $\| \bl \epsilon_i \|$ versus $i$.

\begin{figure}
\begin{center}
\includegraphics[width=0.5\textwidth]{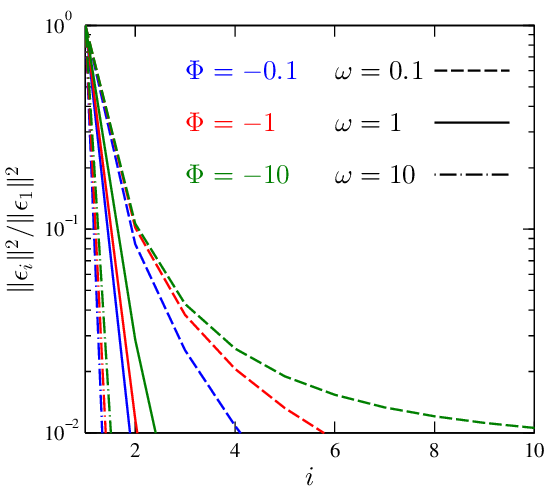}
\caption{Variation of the magnitude of the leading order terms in Eq.~\eqref{F_guess} $\| \bl \epsilon_i \|$ versus $i$ for different values of $\omega$ and $\Phi$. 
As $|\Phi|/\omega$ increases, $\| \bl \epsilon_i \|$ decays slower and the effect of higher order terms on $\lambda$ becomes more pronounced (colors online).}
\label{fig:epsilon_decay}
\end{center}
\end{figure}

The first three terms of the asymptotic expansion are derived and provided in Table \ref{table:f}.
As the order of expansion $n$ increases, additional terms appear in the lower order terms as a result of the interaction between higher order terms. 
All these additional terms are produced by the nonlinear term in Eq.~\eqref{F_eq} that turns into a convolution in the Fourier space.
Further analysis of these terms shows that the decay rate of $\| \bl \epsilon_i \|$ depends primarily on $|\Phi|/\omega$ in the one-dimensional unimodal excitation setting. 
In general $\|\bl \epsilon_i\| \propto (|\Phi|/\omega)^i$ if $|\Phi|/\omega < 1$ and in the worst-case scenario $\|\bl \epsilon_1\| \approx \|\bl \epsilon_2\| \approx \cdots \approx \| \bl \epsilon_n \|$.
In the latter case, the asymptotic form in Eq.~\eqref{F_guess} will not converge. 
To show this behavior in practice, $\bl \epsilon_i$ is derived by continuing Table \ref{table:f} beyond $n=3$.
$\bl \epsilon_i$ and $\lambda$ are then calculated iteratively for several values of $\omega$ and $\Phi$.
The decay rate of $\| \bl\epsilon_i \|$ versus $i$ is then computed and shown in Figure \ref{fig:epsilon_decay}.
This figure confirms that $\| \bl \epsilon_i \|$ may not decay monotonically if $|\Phi|/\omega \gg 1$, as is the case of $\Phi=-10$ and $\omega=0.1$. 
Non-converging $\| \bl \epsilon_i(\omega,\Phi) \|$ is particularly observed where the prediction of  Eq.~\eqref{lambda} disagrees with the reference numerical result.

\begin{table}[H]
\centering
\begin{tabular}{ccccc}
\hline
$n$ & $i$ & $\bl \epsilon_i(\omega)$ & $n(\lambda^2 + \lambda)$ \\
\hline
\hline
1 & 1 & $\bl G(\omega) \left(1 + 2 \lambda + \hat i \omega \right)^{-1}$ & $-2 \bl \epsilon_1(\omega):\bl \epsilon_1^{\rm T}(-\omega) $ \\
\hline
2 & 1 & $\left(\bl G(\omega) - 2\bl \epsilon_2(\omega) \cdot \bl \epsilon_1(-\omega) \right)\left(1 + 2 \lambda + \hat i \omega \right)^{-1}$ &  \\
2 & 2 & $- \bl \epsilon_1(\omega)\cdot \bl \epsilon_1(\omega) \left(1 + 2 \lambda + 2 \hat i \omega \right)^{-1}$ & $-2\sum_{j=1}^n \bl\epsilon_j(\omega):\bl\epsilon_j^{\rm T}(-\omega)$\\
\hline
3 & 1 & $\left(\bl G(\omega) - 2\bl \epsilon_2(\omega)\cdot\bl \epsilon_1(-\omega) -2 \bl \epsilon_3(\omega)\cdot  \bl \epsilon_2(-\omega) \right) \left(1 + 2 \lambda + \hat i \omega \right)^{-1}$ & \\
3 & 2 & $\left(-2\bl \epsilon_3(\omega)\cdot \bl \epsilon_1(-\omega) - \bl \epsilon_1(\omega)\cdot\bl \epsilon_1(\omega) \right) \left(1 + 2 \lambda + 2 \hat i \omega \right)^{-1}$ & \\
3 & 3 & $-2\bl \epsilon_2(\omega)\cdot \bl \epsilon_1(\omega) \left(1 + 2 \lambda + 3 \hat i \omega \right)^{-1}$ & $-2\sum_{j=1}^n \bl\epsilon_j(\omega):\bl\epsilon_j^{\rm T}(-\omega)$ \\
\hline
\end{tabular}
\caption{The leading order terms in the asymptotic solution of Eq.~\eqref{F_eq}.
The solution of the first order expansion ($n=1$) for $\lambda$ is already provided in Eq.~\eqref{lambda}.
For $n>1$, $\bl \epsilon_i$ and $\lambda$ must be calculated iteratively.
For arbitrary tensors $\bl A$ and $\bl B$, $(\bl A \cdot \bl B)_{ij}$ is defined as $\frac{1}{2} \left( A_{ik}B_{kj} + A_{jk}B_{ki} \right)$.}
\label{table:f}
\end{table}

Neglecting the higher order terms accounts for only some of the observed discrepancies.
For a certain combination of $\omega$ and $\Phi$, including higher order terms does not produce a better estimate for $\lambda$.
In these cases, $\| \bl \epsilon_i \|$ may not even decay for $i>1$, suggesting that the assumed form in Eq.~\eqref{F_guess} is incomplete.
This form assumes that excitation at $\omega$ would generate a solution oscillating at $\omega, 2\omega, 3\omega, \cdots$, producing only super-harmonics. 
The numerical simulation shows, however, that the solution may oscillate at lower frequencies, producing sub-harmonics.
A closer examination shows that the amplitude of sub-harmonics can far exceed that of super-harmonics.
It is these sub-harmonics that cause highly nonlinear behaviors that are missing in the prediction of the present analysis.
Identifying the nontrivial patterns of these sub-harmonics and thereby accounting for their contribution to $\lambda$ remains as a topic for future studies. 

\bibliographystyle{elsarticle-num}

\end{document}